\documentclass[conference]{IEEEtran}
\IEEEoverridecommandlockouts
\usepackage{cite}
\usepackage{amsmath,amssymb,amsfonts}
\usepackage{graphicx}
\usepackage{textcomp}
\usepackage{xcolor}
\def\BibTeX{{\rm B\kern-.05em{\sc i\kern-.025em b}\kern-.08em
    T\kern-.1667em\lower.7ex\hbox{E}\kern-.125emX}}
    
\usepackage{graphicx}
\usepackage{bbm}
\usepackage{bm}
\usepackage{amsmath,amsfonts}
\usepackage{amsthm}
\newtheorem{definition}{Definition}
\usepackage{slashed}
\usepackage{booktabs}
\usepackage{subfig}
\usepackage{algorithm}
\usepackage[noend]{algpseudocode}

\usepackage{color}
\definecolor{XFcolor}{RGB}{209,186,116}
\definecolor{XFcolorb}{RGB}{25,202,173}
\usepackage[braket]{qcircuit}

\begin{document}

\title{Quantum Algorithm for Maximum Biclique Problem
\thanks{This research was funded by the Federal Ministry of
Education and Research (BMBF), Germany under the project
LeibnizKILabor with grant No.\,01DD20003.}
}
\author{\IEEEauthorblockN{Xiaofan Li\IEEEauthorrefmark{2},
Prasenjit Mitra\IEEEauthorrefmark{2}, 
Rui Zhou\IEEEauthorrefmark{3},
 and
Wolfgang Nejdl\IEEEauthorrefmark{2}}
\IEEEauthorblockA{
\IEEEauthorrefmark{2}L3S Research Center, Germany. Email: \{xiaofan.li,mitra,nejdl\}@l3s.de\\
\IEEEauthorrefmark{3}
Swinburne University of Technology, Australia.  
Email: rzhou@swin.edu.au
}

}

\maketitle

\begin{abstract}
Identifying a biclique with the maximum number of edges bears considerable implications for numerous fields of study and application, such as detecting anomalies in E-commerce transactions, discerning protein-protein interactions in biological studies, and refining the efficacy of social network recommendation algorithms. However, the inherent NP-hardness of this problem significantly complicates the matter. 
The prohibitive time complexity of existing algorithms is the primary bottleneck constraining the application scenarios. 
Another obstacle resides in the ever-increasing energy requirements for running these algorithms. The escalating power consumption not only exacerbates the economic cost but also poses environmental concerns, making it increasingly difficult to deploy these solutions in applications sustainably.
Aiming to address these challenges, we present an unprecedented exploration of a quantum computing approach to solve the  issues. 
Efficient quantum algorithms, as a crucial future direction for handling NP-hard problems, are presently under intensive investigation. 
Regular advancements in quantum hardware have gradually made quantum computing a more accessible, faster, and cost-effective tool. Its potential has already been proven in practical arenas such as cybersecurity, marking a promising future for this technology. 
However, in the field of quantum algorithms for graph databases, little work has been done due to the challenges presented by the quantum representation of complex graph topologies. 
In this study, we delve into the intricacies of encoding a bipartite graph on a quantum computer. We further design a sub-procedure capable of recognizing whether a given subgraph constitutes a biclique of a given size. Given a bipartite graph with $n$ vertices, we propose a ground-breaking algorithm, dubbed qMBS. This novel methodology can pinpoint a solution within $O^*(2^{\frac{n}{2}})$ iterations of the subprocedure, illustrating a quadratic speed-up in terms of time complexity compared to the state-of-the-art algorithms.
Further expanding the utility of qMBS, we detail two variants tailored for the maximum vertex biclique problem and the maximum balanced biclique problem. To corroborate the practical performance and efficacy of our proposed algorithms, we have conducted proof-of-principle experiments utilizing  advanced quantum simulators available to date.
The important feature of qMBS is its reversible computing manner, which, according to Landauer's Principle, holds substantial promise in dealing with applications with significantly reduced power consumption in the near future. 
\end{abstract}


\section{Introduction}
\textbf{Problem.}
A bipartite graph, represented as $G(L,R,E)$, is structured around two separate and non-overlapping vertex sets, $L$ and $R$, and an edge set $E$ that is a subset of the Cartesian product of $L$ and $R$, or $E \subseteq L \times R$.
A biclique is a particular type of subgraph, which consists of two vertex sets, $A$ and $B$. Here, $A$ is a subset of $L$ ($A \subseteq L$), and $B$ is a subset of $R$ ($B \subseteq R$). The distinguishing feature of a biclique is that every vertex in set $A$ is connected to - or neighbors with - every vertex in set $B$.
Among all such bicliques, a maximum biclique is defined as the one that possesses the greatest number of edges. This essentially refers to the largest complete bipartite subgraph.
The focus of this study is the Maximum Biclique Problem (MBP), the challenge of finding such a maximum biclique within a given bipartite graph. 

\textbf{Significance.}
The concept of a biclique is foundational to an array of applications across diverse fields: 
\begin{itemize}
	\item [(1)] In the realm of E-commerce, the anomaly detection process~\cite{allahbakhsh2013collusion,beutel2013copycatch} often necessitates identifying clusters of customers who collectively purchase a set of products. Such coordinated behaviors frequently flag potential instances of fraudulent product ranking manipulation. 
 Therefore, identifying the maximum biclique could aid in pinpointing the largest group involved in illicit click activities within E-commerce networks, thus curbing fraudulent activities. 
	\item [(2)] In the field of biological studies, protein-protein interactions~\cite{mukhopadhyay2014incorporating,kaloka2019pols,bustamam2020application,dey2020graph,zhou2022comprehensive} are crucial. Researchers strive to uncover groups of human proteins that interact with the same set of viral proteins, such as those belonging to HIV and SARS-CoV-2, the virus that causes COVID-19. 
 Therefore, finding the largest biclique can lead to the discovery of the most significant disease-causing protein group, potentially offering breakthroughs in combating viruses like HIV or COVID-19. 
	\item [(3)] The strategy for social network recommendation systems~\cite{liu2006efficient} often relies on recognizing sets of users who exhibit shared interests, thereby enhancing the efficacy of targeted advertising. 
 Identifying the user group with the highest potential market value for advertising could drastically improve the efficiency and return on investment of marketing strategies.
\end{itemize}

\textbf{Uncharted opportunity.}
The Maximum Biclique Problem (MBP) has been identified as NP-hard~\cite{peeters2003maximum}, and it has been convincingly demonstrated that it is highly challenging, if not impossible, to develop a polynomial time algorithm that boasts a substantial approximation ratio~\cite{ambuhl2011inapproximability,manurangsi2018inapproximability}. The current state-of-the-art solution for the MBP~\cite{lyu2022maximum} has a time complexity of $O(2^n)$.
The prohibitive time complexity of the state-of-the-art is the primary bottleneck constraining the application scenarios.
A promising alternative, however, is offered by quantum computing, an emerging technology set to revolutionize computational paradigms for NP-hard problems in the foreseeable future~\cite{preskill2018quantum}. 
The inherent parallelism and the unique properties of quantum superposition and entanglement provide novel pathways for solving these  problems~\cite{nielsen2010quantum}. Particularly, Quantum algorithms like Shor's algorithm for factorization and Grover's search algorithm showcase potential exponential and quadratic speedups, respectively, over their classical counterparts~\cite{shor1999polynomial,grover1996fast}. 
Recent studies have highlighted that a specialized quantum algorithm is likely to provide a quadratic speed-up in terms of time complexity over a classical algorithm when applied to NP-hard problems~\cite{hao2001quantum,gaitan2014graph,su2016quantum,srinivasan2018efficient}.
The challenge to the actual implementation of quantum algorithms
 is the current stage of quantum hardware, which, as of now, has not yet reached the fault-tolerant quantum computing regime, also known as quantum error correction, necessary for running quantum algorithms at scale~\cite{preskill2018quantum}.
A promising direction to address this challenge is the extensive integration of quantum computing with classical computing using cloud computing platforms~\cite{brown2022quantum}. 
Given the current rate of progress in the field, researchers estimate that a timeline of 5 to 15 years may be plausible for quantum computers to be ready to solve large-scale, real-world database problems~\cite{preskill2018quantum,arute2019quantum}.
However, compiling classical graph problems into the quantum computation model that can run on large-scale Quantum Processing Units (QPUs) is a non-trivial task due to the constraints in qubit  operations~\cite{moll2018quantum}  and quantum representation of complex graph topologies. 
One of the contemporary challenges in the field lies in the development of quantum algorithms for the extant NP-hard problems in graph databases. These algorithms should be viable for execution on prospective large-scale QPUs, while also being amenable to simulation and proof-of-principle experimentation on the limited-scale QPUs that currently prevail in the quantum computing landscape.

 Another crucial advantage of quantum computing resides in an often-overlooked fact of computational theory: energy consumption. As data sizes continue to grow, the energy resources, such as electricity, consumed by a normal computer typically escalate alongside the curve defined by time complexity. 
Normal computers consume energy in the process of computation due to Landauer's Principle~\cite{5392446}. According to this principle, when a computer erases a bit of information, a minimum amount of energy must be dissipated into the environment. Normal computers continually erase information during computation, as their operational logic is based on irreversible logic gates. For example, an AND gate maps two bits of information into one bit, and it is not possible to infer the original two-bit input from the resultant one-bit output.
In contrast, quantum algorithms employ reversible gates for computation, thereby avoiding information erasure. In principle, this makes the process dissipation-free. In practicality, some energy dissipation is still required for system stability and to provide immunity from noise. Nevertheless, quantum computing, when utilized in combination with appropriately designed algorithms, still presents a promising strategy for 
graph computation that is significantly more energy-efficient than normal computing~\cite{nielsen_chuang_2010}.

\textbf{Our approach.}
In this research, 
we focus on addressing the challenge of reducing the time complexity of the existing maximum biclique algorithms when executed on large-scale QPUs.
We develop a novel, reversible quantum algorithm termed as qMBS. This algorithm exhibits a time complexity of $O^*(2^{\frac{n}{2}})$, where $n$ represents the number of vertices.\footnote{For a quantum algorithm, quantum complexity classes apply.}
Our approach utilizes the foundational framework of Grover's search algorithm~\cite{grover1997quantum}, a quantum circuit-based solution designed for unstructured database searches. The crux of Grover's search algorithm lies in an {\it oracle}, a mechanism that {\it identifies} the query item. For our application, this oracle is utilized to (1) ascertain whether a subgraph forms a biclique, and (2) determine the size of the subgraph.
We innovatively design this oracle using reversible computational units, known as quantum gates, to execute these two tasks. Consequently, our comprehensive algorithm, qMBS, achieves a quadratic speed-up over the state-of-the-art in terms of time complexity~\cite{lyu2022maximum} with significantly reducing energy dissipation.

We highlight our principal contributions below.
\begin{itemize}
\item We introduce a versatile design that encodes a bipartite graph into a quantum circuit. This approach is broadly applicable to an array of biclique problems, including the maximum vertex biclique problem and maximum balanced biclique problem.
\item We illustrate the mapping of our problem into this framework by utilizing qMBS, an innovative algorithm developed by adapting the principles of Grover's search. qMBS incorporates a dedicated oracle to ascertain whether a given subgraph is a biclique of a specified size. With a time complexity of $O^*(2^{\frac{n}{2}})$, our approach provides a quadratic speed-up over the state-of-the-art in terms of complexity. 
Further, its inherent reversible computing mechanism promises 
an economical computation manner in the near future.
\item We conduct proof-of-principle experiments utilizing state-of-the-art quantum simulators, validating the practical performance and efficacy of our proposed algorithms.
\end{itemize}


\textbf{Roadmap.}
The remainder of the paper is organized as follows.
Section~\ref{sec:pre} reviews the preliminaries.
Section~\ref{sec:qMBS} introduces our algorithm qMBS and its variants for other biclique problems. 
Section~\ref{sec:exp} conducts experimental studies. 
Related works and conclusion are in Section~\ref{sec:related} and Section~\ref{sec:conclu}.

\section{Preliminaries}\label{sec:pre}

In this section, we will revisit some of the fundamental concepts related to bicliques and provide a succinct introduction to quantum computing, specifically focusing on the computational model of quantum circuits. Subsequently, we will present Grover's search as the fundamental framework that underpins our proposed methodologies.

\subsection{Maximum Biclique Problem}

Our study focuses on an unweighted and undirected bipartite graph, denoted as $G(L,R,E)$. Here, $L$ and $R$ represent two separate sets of vertices, while $E\subseteq L\times R$ signifies the set of edges. The graph's size is characterized by $n = |L| + |R|$ (number of vertices) and $m = |E|$ (number of edges). When referring to a subgraph $C$, we also use $C$ to denote its subset of vertices, for clarity within the given context. Consequently, we denote $L(C) = C\cap L$ and $R(C) = C\cap R$ to express the intersection of subgraph $C$ with vertex sets $L$ and $R$, respectively.
A biclique is a complete bipartite subgraph of $G$: 
\begin{definition}[\textbf{Biclique}]
	Given a bipartite graph $G(L,R,E)$, a biclique $C$ is a subgraph of $G$, $s.t.$ for each pair of $u\in L(C)$ and $v\in R(C)$, 
the edge $(u,v)\in E$ exists.
\end{definition}

\begin{definition}[{\small\textbf{Maximum Biclique Problem (MBP)}}]
	Given a bipartite graph, 
	find a biclique with the maximum 
	edge number. 
\end{definition}
MBP is NP-hard~\cite{peeters2003maximum}, 
and it is difficult to find a polynomial time algorithm 
with a promising approximation ratio~\cite{ambuhl2011inapproximability,manurangsi2018inapproximability}. 
The state-of-the-art~\cite{lyu2022maximum} 
has a time complexity  $O(2^n)$. 
In this work, we propose an algorithm to solve MBP in $O^*(2^{\frac{n}{2}})$. 
\subsection{Quantum Mechanics/Computing}
Quantum mechanics studies how to describe the state of a microscopic system (e.g., an atom), and how such a state evolves over time. 
Quantum computing 
focuses on 
 how to (1) encode a computation problem into the state of a microscopic system; (2) evolve such a state into the final solution state.  
Mathematically, a quantum state is represented as a vector, and the principles governing its evolution are characterized through vector rotations. In the context of this study, we engage with the most basic quantum system, the state of which is referred to as a qubit:
\begin{definition}[\textbf{Qubit}]
	A qubit is a vector with a unit norm in a two-dimensional complex linear space: 
	\begin{equation}
		\ket{q} = \alpha \ket{0} + \beta \ket{1}
	\end{equation}
	Here we use the notation $\ket{\cdot}$ to denote a vector. 
	$\ket{0}$ and $\ket{1}$ are  base vectors of the   space.  
	The complex coefficients $\alpha$ and $\beta$ are called amplitudes, 
	which satisfy $|\alpha|^2 + |\beta|^2 = 1$. 
\end{definition}
A qubit sets itself apart from a traditional bit, whose state is strictly either 0 or 1. In contrast, the state of a qubit is a superposition, $\alpha \ket{0} + \beta \ket{1}$, which is neither strictly $\ket{0}$ nor $\ket{1}$. This can be visualized as a composite vector distinct from both of the base vectors.
Owing to the continuous nature of the coefficients, the amount of information that a qubit can theoretically hold is limitless, offering an intuitive understanding of the superior potential of quantum computing compared to conventional computing. However, this information cannot be directly accessed because a measurement of the superposition state $\ket{q}$ will result in a random \textit{collapse} to the base state $\ket{0}$ with a probability of $|\alpha|^2$, or to $\ket{1}$ with a probability of $|\beta|^2$.
Therefore, it is crucial to devise skillfully designed quantum algorithms to manage the information encoded within a qubit.

When considering a system comprised of $n$ qubits, the state of the system is expressed using a tensor product. For instance, the state of a two-qubit system is represented as follows:
%
\begin{equation}
	\begin{aligned}
		\ket{q_{comp}} &= \ket{q_1} \ket{q_2}\\
		&= (\alpha_1 \ket{0} + \beta_1 \ket{1})(\alpha_2 \ket{0} + \beta_2 \ket{1}) \\
		&= \alpha_1\alpha_2\ket{00} + \alpha_1\beta_2\ket{01}
		+ \alpha_2\beta_1\ket{10} + \beta_1\beta_2\ket{11}
	\end{aligned}
\end{equation} 
We will use $\ket{ij}$ to denote $\ket{i}\ket{j}$.
State evolves  by vector rotation, which is described by matrix multiplication:
\begin{equation}
	\ket{q_{initial}}\xrightarrow[]{\mbox{over time}}\ket{q_{final}} = U\ket{q_{initial}}
\end{equation}
$U$ is a unitary matrix satisfying $U^{\dag}U = I$, 
where ${\dag}$ is conjugate transpose and $I$ is the identity matrix. 
E.g.,
a matrix $X$ evolves a qubit by turning $\ket{0}$ into $\ket{1}$ and turning $\ket{1}$ into $\ket{0}$: 
\begin{equation}\label{eq:X}
	X\ket{q} = \alpha X\ket{0} + \beta X\ket{1} = \alpha \ket{1} + \beta \ket{0}
\end{equation}
Another matrix utilized in our study is the Hadamard matrix $H$. This matrix transforms $\ket{0}$ into an equal superposition state $(\ket{0} + \ket{1})/\sqrt{2}$ and transforms $\ket{1}$ into $(\ket{0} - \ket{1})/\sqrt{2}$. Typically, the $H$ matrix is used for initial state preparation. By operating on the equal superposition state, a single quantum operation can potentially act on all possible states concurrently, harnessing the power of quantum parallelism.

 If we explicitly write the two base vectors as $\ket{0} = [1,0]^T, \ket{1} = [0,1]^T$, then it can be verified that the matrix $X$ and $H$ can be written as 
 \begin{equation}
 	X = 
 	\begin{bmatrix}
 		0 & 1 \\
 		1 & 0
 	\end{bmatrix},\hspace{.4cm}
 	H = \frac{\sqrt 2}{2}
 	\begin{bmatrix}
 		1 & 1 \\
 		1 & -1
 	\end{bmatrix}
 \end{equation}


\subsection{Computation Model: Quantum Circuit}

A quantum circuit provides a visual representation of state evolution, consisting of quantum wires (representing qubits) and quantum gates (representing matrices). Figure~\ref{fig:xgate} displays a basic quantum circuit where the qubit $\ket{0}$ transitions to $\ket{1}$ after the application of the $X$ matrix, i.e., a quantum $X$ gate. The progression towards the right signifies the flow of time.
The $X$ gate bears a resemblance to a logical {NOT} gate as it effectively flips a bit. However, the crucial difference lies in their target state: while the logical {NOT} gate operates on a definite state (either $0$ or $1$), the quantum $X$ gate acts on a superposition state (refer to Eq.~\ref{eq:X}). This capability for parallel operation is a key reason behind the remarkable speed of quantum algorithms.


\begin{figure}[t]
\[ \Qcircuit @C=1em @R=.7em {
      \lstick{\ket{0}}& \gate{X} & \rstick{\ket{1}}\qw
}\]
\caption{A toy quantum circuit}
\label{fig:xgate}
\end{figure}

\begin{figure}[t]
\[ \Qcircuit @C=1em @R=.7em {
    \hspace{-2.2cm} \mbox{control} & \ctrl{1} & \qw & && \ctrl{1} & \qw \\
     \hspace{-2.2cm}\mbox{target} & \gate{X} & \qw &\ustick{=} && \targ & \qw 
}\]
\caption{Two representations of a CNOT gate}
\label{fig:cnot}
\end{figure}
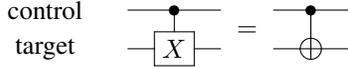

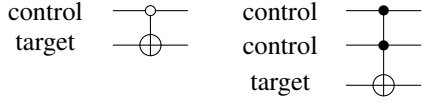
\begin{figure}[t]
\[ \Qcircuit @C=1em @R=.7em {
    \hspace{-1.8cm} \mbox{control} & \ctrlo{1} & \qw & &&\hspace{10pt}\mbox{control} &&&&\ctrl{1} & \qw \\
     \hspace{-1.8cm}\mbox{target} & \targ & \qw & &&  \hspace{10pt}\mbox{control} &&&&\ctrl{1} & \qw \\
     &&&&&\hspace{10pt}\mbox{target} &&&&\targ & \qw
}\]
\caption{Other types of controlled-gate}
\label{control}
\end{figure}

\begin{figure*}[t]
\vspace{5pt}

	\centering
	{\includegraphics[width=18cm]{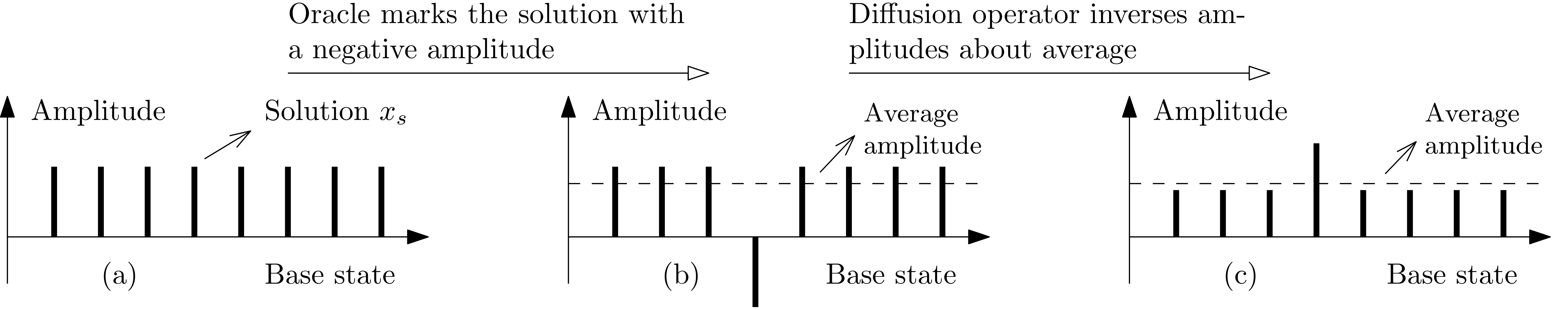}}

	\captionsetup{justification=centering}
	\caption{Illustration of the Grover's search with $n = 3$}\label{grover}
\end{figure*}

Another quantum gate utilized in this study is the controlled-$X$ gate, also known as the CNOT gate (see Figure~\ref{fig:cnot}). The control qubit, designated by a solid circle on a quantum wire, dictates the operation on the target qubit. If the control qubit is in state $\ket{1}$, the $X$ gate operates on the target qubit; if the control qubit is in state $\ket{0}$, the target qubit remains unaltered.
For convenience, we will subsequently represent the $X$ gate with a circle encompassing a cross. Alternatively, the target can be flipped when the control qubit is in state $\ket{0}$, denoted by a hollow circle.
Controlled gates can be further specified with additional control qubits. In this case, the target qubit will only be operated on when all control qubits align with their respective base states. A CNOT gate with $k$ control qubits is denoted as a C$^k$NOT gate. Examples are provided in Figure~\ref{control}.

\subsection{Grover's Search}

The framework of our algorithm   is  Grover's search, 
which was initially designed for unstructured database search: 
\begin{definition}[\textbf{Unstructured Database Search}]
	Given $\mathcal X = \{0, 1, ..., 2^{n}-1\}$ to be a set of  $2^{n}$ integers, 
 a function  $f: \mathcal X\rightarrow \{0,1\}$ satisfies that  
 there exists a unique $x_s\in \mathcal X$, s.t. $f(x_s) = 1$, whereas  for all the other $x\in \mathcal X$ and $x \neq x_s, f(x) = 0$. 
The problem is to find the   $x_s$. 
\end{definition}


Every integer $x\in\mathcal X$ can be represented as an $n$-bit string. Consequently, we can express $x$ as the tensor product of $n$ qubit base states. For instance, the number $3$ can be written as $\ket{0...011}$, or more succinctly, $\ket{3}$.
The fundamental strategy is to operate concurrently on all $2^n$ base states (integers) through superposition. The process iteratively increases the amplitude of the solution state $\ket{x_s}$ until it substantially exceeds the amplitudes of other base states. 
The algorithm is presented as Algorithm~\ref{algo:grover}.

\begin{algorithm}[tttt]
        \caption{Grover's Search Algorithm}
        \begin{algorithmic}[1] 
            \Require A set of integers $\mathcal X = \{0, 1, ..., 2^{n}-1\}$, the discriminant function $f: \mathcal X\rightarrow \{0,1\}$; 
            \Ensure The integer $x_s$ that satisfies $f(x_s) = 1$;
            \State Prepare 
	an equal superposition state $\frac{1}{\sqrt{2^n}}\sum_{i = 0}^{2^n-1}\ket{i}$; 
            \State Use a black box to flip the  amplitude sign of the solution base state $\ket{x_s}$, i.e., from 
	$+\frac{1}{\sqrt{2^n}}\ket{x_s}$ to $-\frac{1}{\sqrt{2^n}}\ket{x_s}$; 
            \State Use a diffusion operator to inverse the amplitude of each base state about the  average of  all the amplitudes; 
			\State Repeat Step 2\&3 for $\lfloor\frac{\pi}{4}\sqrt{2^n}\rfloor$ times, then measure the final state;
			\State Output the binary string read from the final state as $x_s$; 
        \end{algorithmic}
    \label{algo:grover} 
   
    \end{algorithm}


\textbf{Explanation:}
\begin{itemize}
	\item [1.] The equal superposition state is prepared by using $n$ $H$ gates to act on 
	$n$ initial state $\ket{0}$s: 
	\begin{equation*}
		\underbrace{(H\ket{0})(H\ket{0})...(H\ket{0})}_{n} = \frac{1}{\sqrt{2^n}}\sum_{i = 0}^{2^n-1}\ket{i}
	\end{equation*}
	
	The result is illustrated as Figure~\ref{grover}a. 
	In this figure, We utilize a bar graph to represent a superposition state, where the x-axis corresponds to different base states (basis vectors), and the y-axis indicates the amplitude of each base state. In this illustration, we take $n=3$ as an example, resulting in a total of $2^3=8$ base states depicted on the graph. As this superposition state is an equal superposition, the amplitudes of all eight base states are of identical height. 
	
\item[2.] A critical component, referred to as an oracle (the black box), is fundamental to Grover's search. It essentially \textit{recognizes} the solution base state $\ket{x_s}$. The resulting state after this step is displayed in Figure~\ref{grover}b. 
Here we observe that the amplitude of the solution base state has been flipped below the x-axis; in other words, its amplitude has been multiplied by a negative sign. The amplitudes of the other non-solution base states remain unchanged. If we compute the average of the amplitudes of all base states, due to the presence of a negative amplitude, the average is slightly less than the amplitudes of the non-solution base states. We mark this average with a dashed line and as can be seen, this average dashed line is slightly lower than the amplitudes of the non-solution base states.

\item[3.] We denote any arbitrary amplitude as $\alpha$ and the average amplitude as $\overline\alpha$. The diffusion operator
transforms any amplitude $\alpha$ into $2\overline \alpha - \alpha$, effecting an inversion about the average. Please refer to Figure~\ref{grover}c. 
Compared to Figure~\ref{grover}a, we observe that the amplitudes of all non-solution base states have decreased, whereas only the amplitude of the solution base state has increased. The actual effect of Line 2 and Line 3 is to suppress the amplitudes of the non-solution base states while amplifying the amplitude of the solution base state.
\item[4.] Steps 2 and 3 cause an increment in the amplitude of the solution state by $O(1/\sqrt{2^n})$. After approximately $\lfloor\frac{\pi}{4}\sqrt{2^n}\rfloor$ iterations, the solution amplitude will be near 1. Consequently, after measurement, the superposition will collapse into the solution state.	
\end{itemize}


Note that if there are $M$ solutions, then finding one solution requires only $\lfloor\frac{\pi}{4}\sqrt{2^n/M}\rfloor$ iterations. If $M$ is unknown, the Quantum Counting algorithm~\cite{brassard1998quantum} can estimate $M$'s value. Intuitively, given a bipartite graph with $n$ vertices, there exist a total of $2^n$ subgraphs. Hence, the search for the maximum biclique can be equated to identifying a solution among the $2^n$ subgraphs. The crucial aspect is to devise a dedicated oracle that can recognize the desired state and flip its amplitude sign. We subsequently demonstrate the construction of such an innovative oracle using a quantum circuit. As the diffusion operator is a universal aspect across various problems,  owing to space constraints, we will not delve into its details.
We summarize the notations utilized throughout this paper in Table~\ref{notations}. 


\begin{table}
\renewcommand\arraystretch{1.2}
\centering
\caption{Notations}
\begin{tabular}{p{40pt}p{190pt}}
\toprule
Notation  & Meaning \\
\midrule
 $G(L,R,E)$ & graph $G$ with vertex set $L$, $R$ and edge set $E$  \\
  $G'(L,R,E')$ & virtual graph with the virtual edge set $E' = L\times R$\\
 $C$ & subgraph or the vertex set of a subgraph\\
 $\ket{q}$ & quantum state, i.e., a complex vector with norm 1\\
 $X$  & quantum not gate\\
 $H$ & Hadamard gate\\
 C$^k$NOT & control-NOT gate with $k$ control qubits\\
 $\oplus$ & XOR logic/modulo two addition\\
 $\bigwedge$ & AND logic\\
 $U, U^{-1}, U^{\dag}$ & unitary matrix, its inverse and  conjugate transpose\\
\bottomrule
\end{tabular}
\label{notations}
\end{table}

\section{A Quantum Algorithm for MBP}\label{sec:qMBS}
Given a bipartite graph $G(L,R,E)$ with $n$ vertices and $m$ edges,
the problem is to find one vertex subset  that is a biclique with the maximum size among $2^n$ subsets.   
We encode $n$ vertices by $n$ qubits using one-hot encoding, i.e., 
using the binary digit $1$ or $0$ to represent whether a vertex is present or absent. 
Then  any $n$-qubit base state can be interpreted as a vertex set. 
Our proposed Quantum Maximum Biclique Search (qMBS)
uses Grover's search with an oracle to search for a biclique with a given size $k\in [1,m]$, 
and uses a binary search to find the maximum $k_M$. 

In this section, we  design the oracle by partitioning it into two parts:
\begin{itemize}
	\item Part I checks whether a base state is a biclique; 
	\item Part II checks whether a base state has a given size $k$. 
\end{itemize}
For better illustration, 
we use the graph in Figure~\ref{bic1} as an example
 thereafter, 
 where 
$L = \{v_1,v_2\}, R = \{u_1,u_2\}, E = \{e_1,e_2,e_3\}$. 
Given a vertex set that is interpreted as a base state $\ket{v_1v_2u_1u_2}$ 
(e.g., $\{v_1,u_2\}$ is represented by $\ket{1001}$ or $\ket{9}$),
the first task is  to determine whether it is a biclique by a quantum circuit. 
We  can get some intuitions by introducing the virtual graph $G'(L,R,E')$, where $E' = L\times R$. 
The virtual graph $G'$ uses $|L| |R|$ virtual edges to connect all the pairs of vertices between $L$ and $R$ (Figure~\ref{bic2}). 
For a base state $\ket{x}$, where $x\in[0,2^n-1]$, 
a necessary and sufficient condition of it being a biclique is: 
if the virtual subgraph induced by $\ket{x}$  contains a virtual edge $e_k'$, 
then the corresponding real edge $e_k$ must exist in the real graph. 
i.e., 
$e_k$ and $e_k'$ must be both present or both absent:   
 \begin{equation}\label{eq:xor}
 	\ket{x}\mbox{ is a biclique}\Longleftrightarrow\bigwedge\nolimits_{k=1}^{|L|  |R|}\overline{(e_k\oplus e_k')}=1
 \end{equation} 
Here $\oplus$ is the XOR logic  (modulo two addition).  

\subsection{Oracle Part I: Biclique Checking}\label{subsec:bic}

\begin{figure}[t]
	\centering
	\subfloat[Real graph $G$\label{bic1}]{\includegraphics[width=3.5cm]{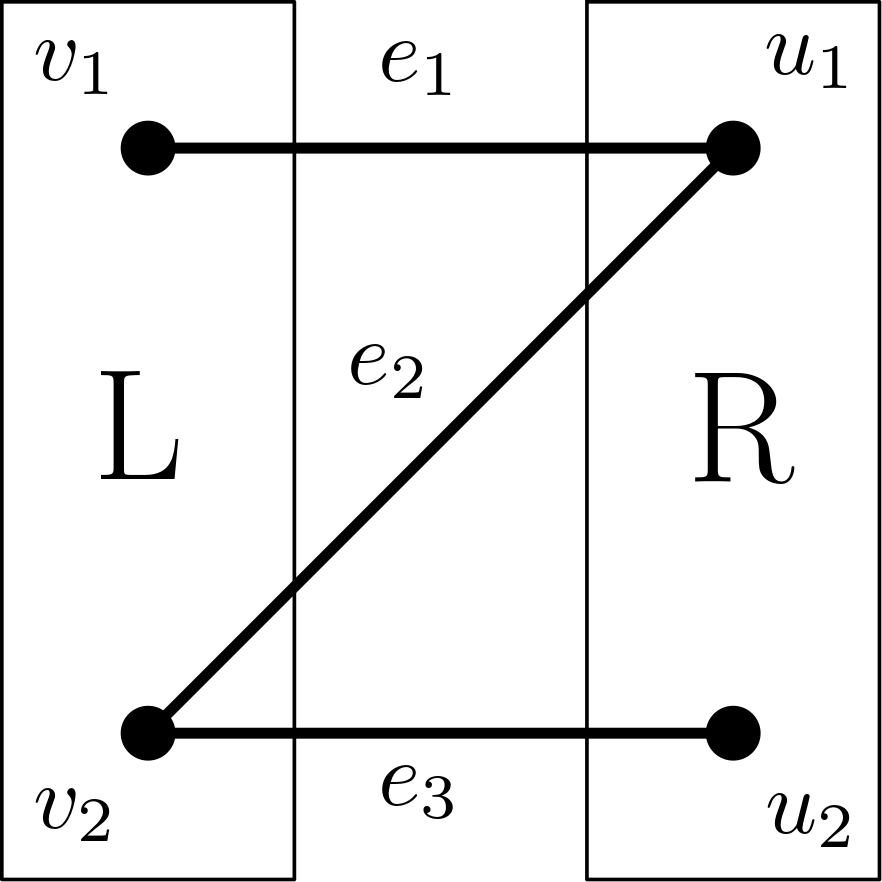}}
	\hspace{25pt}
	\subfloat[Virtual Graph $G'$\label{bic2}]{\includegraphics[width=3.5cm]{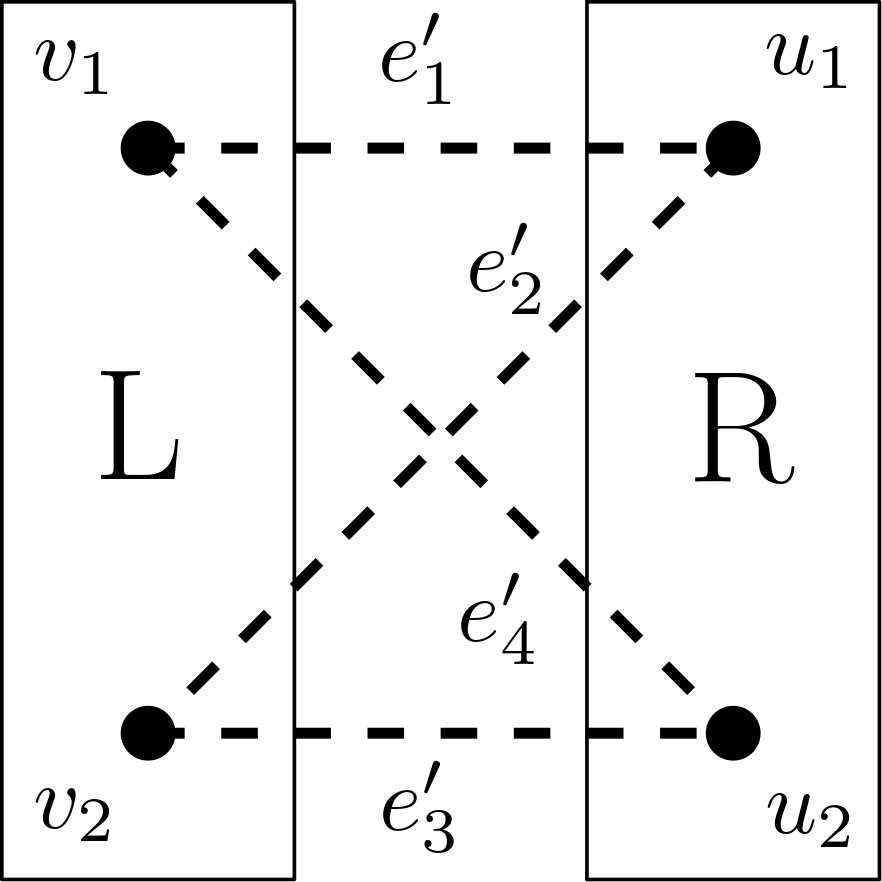}}

	\captionsetup{justification=centering}
	\caption{Example graph}\label{fig:bic}
\end{figure}

\begin{figure}[t]
\[
\Qcircuit @C=.45em @R=.45em @! {
& & &\mbox{real edges} & & & & &\hspace{-0.3cm}\mbox{virtual edges} & & & & &\hspace{0.2cm}\mbox{biclique check} & & & \\
&\lstick{\ket{v_1}} & \ctrl{2} & \qw & \qw & \qw & \ctrl{2}  & \qw & \qw & \ctrl{3}& \qw & \qw & \qw & \qw & \qw & \qw & \qw\\
&\lstick{\ket{v_2}} & \qw & \ctrl{1} & \ctrl{2} & \qw & \qw  & \ctrl{1} & \ctrl{2} & \qw & \qw & \qw & \qw & \qw & \qw & \qw & \qw\\
&\lstick{\ket{u_1}} & \ctrl{2} & \ctrl{3} & \qw & \qw & \ctrl{6}    & \ctrl{7} & \qw & \qw & \qw & \qw & \qw & \qw & \qw & \qw & \qw\\
&\lstick{\ket{u_2}} & \qw & \qw & \ctrl{3}& \qw & \qw  & \qw & \ctrl{7} & \ctrl{8} & \qw & \qw & \qw & \qw & \qw & \qw & \qw\\
&\lstick{\ket{e_1}}  & \targ & \qw & \qw & \qw & \qw   & \qw & \qw & \qw  & \qw & \ctrl{4} & \qw & \qw & \qw & \qw & \qw\\
&\lstick{\ket{e_2}} & \qw & \targ & \qw & \qw & \qw   & \qw & \qw & \qw & \qw  & \qw & \ctrl{4} & \qw & \qw & \qw & \qw\\
&\lstick{\ket{e_3}} & \qw & \qw & \targ & \qw & \qw   & \qw & \qw & \qw & \qw & \qw & \qw & \ctrl{4} & \qw & \qw & \qw\\
&\lstick{\ket{e_4}} & \qw & \qw & \qw & \qw & \qw   & \qw & \qw & \qw & \qw & \qw & \qw & \qw & \ctrl{4} & \qw & \qw\\
&\lstick{\ket{e_1'}} & \qw & \qw & \qw & \qw &\targ   & \qw & \qw & \qw & \qw & \targ & \qw & \qw & \qw & \ctrlo{1} & \qw\\
&\lstick{\ket{e_2'}} & \qw & \qw & \qw & \qw & \qw  & \targ & \qw & \qw & \qw & \qw & \targ & \qw & \qw & \ctrlo{1} & \qw\\
&\lstick{\ket{e_3'}} & \qw & \qw & \qw & \qw & \qw   & \qw & \targ & \qw & \qw & \qw & \qw & \targ & \qw & \ctrlo{1} & \qw\\
&\lstick{\ket{e_4'}} & \qw & \qw & \qw & \qw & \qw   & \qw & \qw & \targ & \qw & \qw  & \qw& \qw & \targ & \ctrlo{1} & \qw\\
&\lstick{\ket{bic}} & \qw & \qw & \qw & \qw & \qw   & \qw & \qw & \qw & \qw & \qw & \qw & \qw & \qw & \targ & \qw
\gategroup{2}{3}{8}{5}{.7em}{--}
\gategroup{2}{7}{13}{10}{.7em}{--}
\gategroup{6}{12}{14}{16}{.7em}{--}
}
\]
\caption{Biclique check quantum circuit}
\label{complt}
\end{figure}
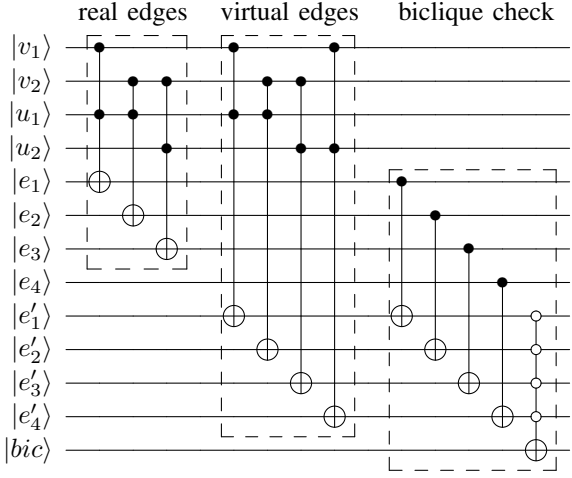

\textbf{Real and virtual graph encoding.}
Figure~\ref{complt} shows the example quantum circuit. 
The $n$ vertices are represented by $n$ qubits $\{\ket{v_i},\ket{u_i}\}$. 
We further use $|L| |R|$ auxiliary qubits $\{\ket{e_i}\}$ to represent real edges and $|L| |R|$ auxiliary qubits $\{\ket{e_i'}\}$ for virtual edges. 
An auxiliary indicator $\ket{bic}$  records the checking result. 
All the auxiliaries  (edges and the indicator) are initially set to be $\ket{0}$. 
Note that although a real edge $e_4$ does not exist in the real graph~\ref{bic1}, we still introduce the qubit $\ket{e_4}$ because we  use $\ket{e_4}\equiv \ket{0}$ to mark its absence,  
and it will be used to compare with $\ket{e_4'}$. 
For any 
real edge $e_k\in E$ that connects two vertices $v_i$ and $u_j$, 
we use a C$^2$NOT gate to connect $\ket{v_i},\ket{u_j}$ and $\ket{e_k}$ 
with $\ket{e_k}$ being the target. 
Please refer to the dashed box with the title \textit{real edges} in Figure~\ref{complt}.
Given a base state (vertex set) $\ket{x}$ with $x\in [0,2^n-1]$, 
these C$^2$NOT gates actually activate all the real edge qubits induced by $\ket{x}$ to be $\ket{1}$.  
For example, given $\ket{x} = \ket{5} = \ket{0101}$, the real edge $\ket{e_3}$ will be activated to be $\ket{1}$, whereas $\ket{e_1},\ket{e_2}$ and $\ket{e_4}$ are all kept in $\ket{0}$. 
Similarly, we construct all the virtual edges using C$^2$NOT gates according to the virtual graph. 
Please refer to the dashed box with the title \textit{virtual edges} in Figure~\ref{complt}. 
By now, we have encoded the real graph and the virtual graph  into the circuit. 
Given a base state $\ket{x}$, the induced real 
and virtual edge qubits will be activated to be $\ket{1}$.

\textbf{Real and virtual edge comparison.}
The remaining work is to compare each $\ket{e_k}$ with the corresponding $\ket{e_k'}$ 
by Eq.~\ref{eq:xor}. 
The XOR logic is implemented by a CNOT gate, 
because $CNOT\ket{e_k}\ket{e_k'} = \ket{e_k}\ket{e_k \oplus e_k'}$,  
where the XOR result is stored into the virtual edge  qubit. 
After using $|L||R|$ CNOT gates to act on all the pairs of real and virtual edges, the virtual edge set $\{\ket{e_k'}\}$   transforms to $\{\ket{e_k\oplus e_k'}\}$. 
The last two steps are first implementing the NOT logic to  all the $\ket{{e_k\oplus e_k'}}$s, 
then using the AND logic to combine them   and store the result into the $\ket{bic}$. 
These two steps are accomplished  by a C$^{|L||R|}$NOT gate with hollow circles. 
Please refer to the dashed box with the title \textit{biclique check} in Figure~\ref{complt}. 
According to Eq.~\ref{eq:xor}, 
if a base state  $\ket{x}$  is a biclique, 
the indicator $\ket{bic}$ will be flipped from $\ket{0}$ to $\ket{1}$. 
Note that if $\ket{x}$  contains only vertices in $L$ or  $R$, we still mark it as a biclique with size 0. 

\textbf{Example results.}
If we represent a physical state as $\ket{v_1v_2u_1u_2}\ket{bic}$, then the input to the algorithm, that is, the initial state, is: 
\begin{equation}
	\ket{state} = \ket{0000}\mathbf{\ket{0}}
\end{equation}

The first step of the algorithm is to use Hadamard gates to prepare an equation superposition state. After being acted by four Hadamard gates, the initial state has been evolved into: 
\begin{equation}
	\begin{aligned}
		\ket{state} = &\frac{1}{4}\big [ \ket{0000}+\ket{0001}+\ket{0010}+\ket{0011}\\
		&+\ket{0100}+\ket{0101}+\ket{0110}+\ket{0111}\\
		&+\ket{1000}+\ket{1001} + \ket{1010}+\ket{1011}\\
		&+\ket{1100}+\ket{1101}+\ket{1110}+\ket{1111}\big ]\mathbf{\ket{0}}
	\end{aligned}
\end{equation}
The appearance of the $1/4$ coefficient here is because we require the entire state, viewed as a vector, to be normalized.
After being processed by the biclique check circuit, the vertex qubits $\ket{v_1v_2u_1u_2}$ should be entangled with the biclique check qubit $\ket{bic}$, which means, $\ket{bic}$ should classify all the base states (subgraphs) into two categories: bicliques (marked by $\ket{bic}=1$) or non-bicliques (marked by $\ket{bic}=0$). 
Then, the result state is shown as: 


\begin{equation}
	\begin{aligned}
		\ket{state}=&\frac{1}{4}\big [ (\ket{0000}+\ket{1000}+\ket{0100}+\ket{0010}+\ket{0001}\\
		&+\ket{1010}+\ket{0101}+\ket{0110}+\ket{1110}+\ket{0111})\mathbf{\ket{1}}\\
		&+ (\ket{1100}+\ket{0011}+\ket{1001}\\
		&+\ket{1101}+\ket{1011}+\ket{1111})\mathbf{\ket{0}}
	\big ]
	\end{aligned}
\end{equation}

Note that in this case, 
 we treat both the empty set and each single vertex as a biclique. This does not affect our search for the maximum biclique, because by definition, both types of bicliques have a size of 0. In the next step when checking the biclique size, the algorithm will automatically disregard bicliques of size 0.
 


\textbf{Summary:}
The biclique check circuit first encodes  the real   and    virtual edges, 
then uses the XOR logic to compare each pair of 
them, 
finally uses the AND logic to store the check result into the indicator. 
In this circuit, the  qubit number is 
\begin{equation}
	n+|L||R| + |L||R| + 1 = O(n^2)
\end{equation}
The number of CNOT gates is 
\begin{equation}
	m+|L||R|+|L||R|+1 = O(n^2)
\end{equation}

\subsection{Oracle Part II: Edge Counting}\label{subsec:ec}
The next task 
is to determine the sizes of the biclique state $\ket{x}$s. 
Given a biclique $C$,  
the idea is to count the vertex number in $L(C)$,  $R(C)$, and then multiply the two numbers. 

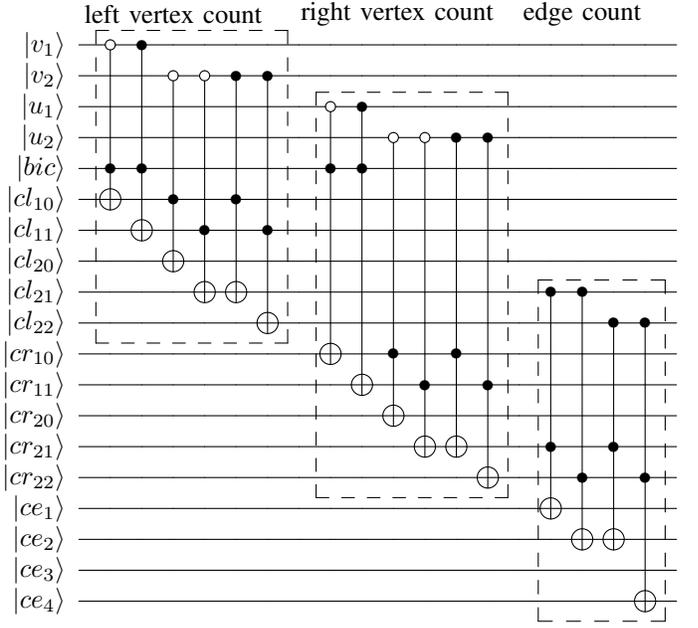
\begin{figure}[t]
\[
\Qcircuit @C=.37em @R=.35em @! {
& & & & \mbox{left vertex count}& & & & & & &\hspace{0.1cm}\mbox{right vertex count} & & & & & &\mbox{edge count} & &  \\
&\lstick{\ket{v_1}} & \ctrlo{4} & \ctrl{4} & \qw & \qw & \qw & \qw & \qw & \qw & \qw & \qw & \qw & \qw & \qw & \qw & \qw & \qw & \qw & \qw & \qw\\
&\lstick{\ket{v_2}} & \qw & \qw & \ctrlo{4} & \ctrlo{5} & \ctrl{4} & \ctrl{5} & \qw & \qw & \qw & \qw & \qw & \qw & \qw & \qw & \qw & \qw & \qw & \qw & \qw\\
&\lstick{\ket{u_1}} & \qw & \qw & \qw & \qw  & \qw & \qw & \qw & \ctrlo{2} & \ctrl{2} & \qw & \qw & \qw & \qw & \qw & \qw & \qw & \qw & \qw & \qw\\
&\lstick{\ket{u_2}} & \qw & \qw & \qw & \qw  & \qw & \qw & \qw & \qw & \qw & \ctrlo{7} & \ctrlo{8} & \ctrl{7} & \ctrl{8} & \qw & \qw & \qw & \qw & \qw & \qw\\
&\lstick{\ket{bic}} & \ctrl{1} & \ctrl{2} & \qw & \qw  & \qw & \qw & \qw & \ctrl{6} & \ctrl{7} & \qw & \qw & \qw & \qw & \qw & \qw & \qw & \qw & \qw & \qw\\
&\lstick{\ket{cl_{10}}} & \targ & \qw & \ctrl{2} & \qw & \ctrl{3} & \qw & \qw & \qw & \qw & \qw & \qw & \qw & \qw & \qw & \qw & \qw & \qw & \qw & \qw\\
&\lstick{\ket{cl_{11}}} & \qw &\targ & \qw & \ctrl{2} & \qw & \ctrl{3} & \qw & \qw & \qw & \qw & \qw & \qw & \qw & \qw & \qw & \qw & \qw & \qw & \qw\\
&\lstick{\ket{cl_{20}}} & \qw & \qw & \targ & \qw & \qw & \qw & \qw & \qw & \qw & \qw & \qw & \qw & \qw & \qw & \qw & \qw & \qw & \qw & \qw\\
&\lstick{\ket{cl_{21}}} & \qw & \qw & \qw & \targ & \targ & \qw & \qw & \qw & \qw & \qw & \qw & \qw  & \qw & \qw & \ctrl{5} & \ctrl{6} & \qw & \qw & \qw\\
&\lstick{\ket{cl_{22}}} & \qw & \qw & \qw & \qw & \qw & \targ & \qw & \qw & \qw & \qw & \qw & \qw  & \qw & \qw & \qw & \qw & \ctrl{4} & \ctrl{5} & \qw\\
&\lstick{\ket{cr_{10}}} & \qw & \qw & \qw & \qw & \qw & \qw & \qw & \targ & \qw & \ctrl{2} & \qw &\ctrl{3}  & \qw & \qw & \qw & \qw & \qw & \qw & \qw\\
&\lstick{\ket{cr_{11}}} & \qw & \qw & \qw & \qw & \qw & \qw & \qw & \qw & \targ & \qw &\ctrl{2} & \qw & \ctrl{3} & \qw & \qw & \qw & \qw & \qw & \qw\\
&\lstick{\ket{cr_{20}}} & \qw & \qw & \qw & \qw & \qw & \qw & \qw & \qw & \qw & \targ & \qw & \qw & \qw & \qw & \qw & \qw & \qw & \qw & \qw \\
&\lstick{\ket{cr_{21}}} & \qw & \qw & \qw & \qw & \qw & \qw & \qw & \qw & \qw & \qw &\targ & \targ & \qw & \qw & \ctrl{2} & \qw & \ctrl{3} & \qw & \qw \\
&\lstick{\ket{cr_{22}}} & \qw & \qw & \qw & \qw & \qw & \qw & \qw & \qw & \qw & \qw & \qw & \qw & \targ & \qw & \qw & \ctrl{2} & \qw & \ctrl{4} & \qw \\
&\lstick{\ket{ce_{1}}}  & \qw & \qw & \qw & \qw & \qw & \qw & \qw & \qw & \qw & \qw & \qw & \qw & \qw & \qw & \targ & \qw & \qw & \qw & \qw \\
&\lstick{\ket{ce_{2}}}  & \qw & \qw & \qw & \qw & \qw & \qw & \qw & \qw & \qw & \qw & \qw & \qw & \qw & \qw & \qw & \targ & \targ & \qw & \qw \\
&\lstick{\ket{ce_{3}}}  & \qw & \qw & \qw & \qw & \qw & \qw & \qw & \qw & \qw & \qw & \qw & \qw & \qw & \qw & \qw & \qw & \qw & \qw & \qw \\
&\lstick{\ket{ce_{4}}}  & \qw & \qw & \qw & \qw & \qw & \qw & \qw & \qw & \qw & \qw & \qw & \qw & \qw & \qw & \qw & \qw & \qw & \targ & \qw 
\gategroup{2}{3}{11}{8}{.7em}{--}
\gategroup{4}{10}{16}{15}{.7em}{--}
\gategroup{10}{17}{20}{20}{.7em}{--}
}
\]
\caption{Edge count / size check quantum circuit}
\label{ecnt}
\end{figure}

\textbf{Vertex count.}
To count the vertices  in $L(C)$, 
we check each vertex in $L$ one by one 
and see whether such a vertex is in $C$. 
We introduce 
a set of auxiliary qubits $\{\ket{cl_{ij}}\}$ 
to record the truth value of the 
proposition that 
\textit{by now we have checked $i$ vertices in $L$ and found that  there are exactly $j$ vertices  contained by $C$, 
where   $i\in [0,|L|], j\in[0,i]$.} 
Updating ${cl_{ij}}$ by checking vertices in $L$ one by one is a dynamic programming procedure:
given $v_{i+1}$ as the truth value whether $(i+1)$th vertex in $L$ is present in $C$, 
the transition equation is 
\begin{equation}\label{eq:trans}
	\begin{aligned}
		&cl_{i+1\mbox{ }j} =  cl_{i\mbox{}j} \wedge\overline {v_{i+1}}\\
		&cl_{i+1\mbox{ }j+1} = cl_{i\mbox{}j} \wedge v_{i+1}
	\end{aligned}
\end{equation}
The initial value $cl_{00}=1$. 
After checking all the  vertices in $L$, 
we get $|L|+1$ values: $\{cl_{|L|\mbox{ }j}\}$, where $j\in [0,|L|]$. 
Among these truth values there is only a single $1$, then the corresponding index $j$ is 
actually $|L(C)|$. 
The circuit design   can be read from the transition equation Eq.~\ref{eq:trans}. 
Each equation is implemented by a C$^2$NOT gate, 
with the L.H.S. being the target qubit. 
Similar to Eq.~\ref{eq:xor}, 
$\overline {v_{i}}$ corresponds to a control qubit marked by a hollow circle. 
Please refer to 
the dashed box with the title \textit{left vertex count} in Figure~\ref{ecnt}.
Here we use $\ket{bic}$ to replace the initial $\ket{cl_{00}}$ 
because we only 
consider
the $\ket{x}$s that are bicliques. 
We count the vertices of $R(C)$ in the same way and store the result into  $\{\ket{cr_{|R|\mbox{ }j}}\}, j\in [0,|R|]$.

\textbf{Multiplication.}
Next 
we have to multiply $|L(C)|$ and $|R(C)|$. 
We introduce a set of auxiliary qubits $\{\ket{ce_k}\}$  to record the result, where $k\in[1,|L||R|]$. 
The multiplication is realized by a mapping 
\begin{equation}
	(\ket{cl_{|L|\mbox{ }i}}, \ket{cr_{|R|\mbox{ }j}}) \mapsto \ket{ce_{i\cdot j}}
\end{equation}
i.e., 
if $\ket{x}$ has $i$ vertices in $L$ and $j$ vertices in $R$, then the edge number will be $i\cdot j$. 
The mapping is realized by a C$^2$NOT gate with the  target being $\ket{ce_{i\cdot j}}$. 
Note that there will be only a single $\ket{1}$ in $\{\ket{ce_k}\}$ since each $\ket{x}$ has a unique size.  
Please refer to the 
dashed box with the title \textit{edge  count} in Figure~\ref{ecnt}.
 
\textbf{Example results.} 
Now we ignore $\ket{bic}$ and use $\ket{v_1v_2u_1u_2}\ket{ce_1 ce_2 ce_3 ce_4}$ to represent the states. 
When $ce_i$ of one particular state is flipped from $0$ to $1$, it means that this state represents a biclique with size $i$. 
Then the states being processed by the edge count circuit has been evolved into: 
\begin{equation}
	\begin{aligned}
		\ket{state} = &\frac{1}{4}\big ( (\ket{1010}+\ket{0101}+\ket{0110})\mathbf{\ket{1000}}\\
		&+ (\ket{1110}+\ket{0111})\mathbf{\ket{0100}}\\
		& + \sum \ket{\mbox{Other States}}\mathbf{\ket{0000}}
		\big )
	\end{aligned}
\end{equation}

We see that each biclique has been marked by a corresponding qubit $\ket{1}$ according to its size, e.g., $\ket{0111}\mathbf{\ket{0100}}$ means the biclique $\{u_1,v_2,u_2\}$ has two edges due to that its $\ket{ce_2} = \ket{1}$.  
Given an arbitrary size $k$, through the implementation of two quantum circuits - biclique check and edge count - we have successfully classified all subgraphs into two categories. The first category comprises bicliques of size $k$, while the second encompasses all remaining subgraphs. Thus far, we have completed the second step in the Grover search process, namely, distinguishing between solutions and non-solutions and subsequently marking them accordingly.

\textbf{Summary.} 
We use a dynamic programming circuit to count 
$|L(C)|$ and $|R(C)|$
for a base state $\ket{x}$, and then use a multiplication mapping to
store the truth value of the 
proposition that \textit{$\ket{x}$ is a size-$k$ biclique} into $\ket{ce_k}$. 
In this circuit, 
the qubit number of $\ket{cl},\ket{cr}$ and $\ket{ce}$
is 
\begin{equation}
	\frac{(2+|L|)(|L|-1)}{2} + \frac{(2+|R|)(|R|-1)}{2} +|L||R| = O(n^2)
\end{equation} 
The number of CNOT gates is 
\begin{equation}
	(1+|L|)|L| + (1+|R|)|R| + |L||R| = O(n^2)
\end{equation}

\begin{figure*}[t]
\[
\Qcircuit @C=.8em @R=-2.1em @! {
& &\mbox{{superposition}} & & &\mbox{oracle to mark the solutions by sign flipping} & & &\mbox{diffusion} & &\hspace{-45pt}\mbox{repeat}  &\mbox{measure}\\
&\lstick{\ket{v_1}} & \gate{H} & \multigate{4}{U_{bic}} & \multigate{8}{U_{size}} & \qw & \multigate{8}{U_{size}^\dag} & \multigate{4}{U_{bic}^\dag} & \multigate{3}{U_{Diff}} & \qw &\hspace{-45pt}... & \meter \\
&\lstick{\ket{v_2}} & \gate{H} & \ghost{U_{bic}} & \ghost{U_{size}} & \qw & \ghost{U_{size}^\dag} & \ghost{U_{bic}^\dag} & \ghost{U_{Diff}} & \qw &\hspace{-45pt}... & \meter\\
&\lstick{\ket{u_1}} & \gate{H} & \ghost{U_{bic}} & \ghost{U_{size}} & \qw & \ghost{U_{size}^\dag} & \ghost{U_{bic}^\dag} & \ghost{U_{Diff}} & \qw &\hspace{-45pt}... & \meter \\
&\lstick{\ket{u_2}} & \gate{H} & \ghost{U_{bic}} & \ghost{U_{size}} & \ustick{\tiny\mbox{}}\qw & \ghost{U_{size}^\dag} & \ghost{U_{bic}^\dag} & \ghost{U_{Diff}} & \qw &\hspace{-45pt}... & \meter \\
&\lstick{\ket{aux}} & \qw & \ghost{U_{bic}} & \ghost{U_{size}} & \ustick{\tiny\mbox{sign flipping}}\qw & \ghost{U_{size}^\dag} & \ghost{U_{bic}^\dag} & \qw & \qw &\hspace{-45pt}...  & \qw\\
&\lstick{\ket{ce_1}} & \qw & \qw & \ghost{U_{size}} & \ctrl{4} & \ghost{U_{size}^\dag} & \qw & \qw & \qw &\hspace{-45pt}... & \qw \\
&\lstick{\ket{ce_2}} & \qw & \qw & \ghost{U_{size}} & \qw & \ghost{U_{size}^\dag} & \qw & \qw & \qw &\hspace{-45pt}... & \qw \\
&\lstick{\ket{ce_3}} & \qw & \qw & \ghost{U_{size}} & \qw & \ghost{U_{size}^\dag} & \qw & \qw & \qw &\hspace{-45pt}... & \qw \\
&\lstick{\ket{ce_4}} & \qw & \qw & \ghost{U_{size}} & \qw  & \ghost{U_{size}^\dag} & \qw & \qw & \qw &\hspace{-45pt}... & \qw \\
&\lstick{\ket{O} = \ket{1}} & \gate{H} & \qw & \qw &  \targ  & \qw & \qw & \qw & \qw &\hspace{-45pt}...  & \qw
\gategroup{2}{3}{5}{3}{.5em}{--}
\gategroup{2}{4}{11}{8}{.9em}{--}
\gategroup{2}{9}{5}{9}{.8em}{--}
\gategroup{7}{6}{11}{6}{1em}{.}
}
\]
\caption{Quantum circuit of qKBS for searching a 1-size  biclique}
\label{full}
\end{figure*}
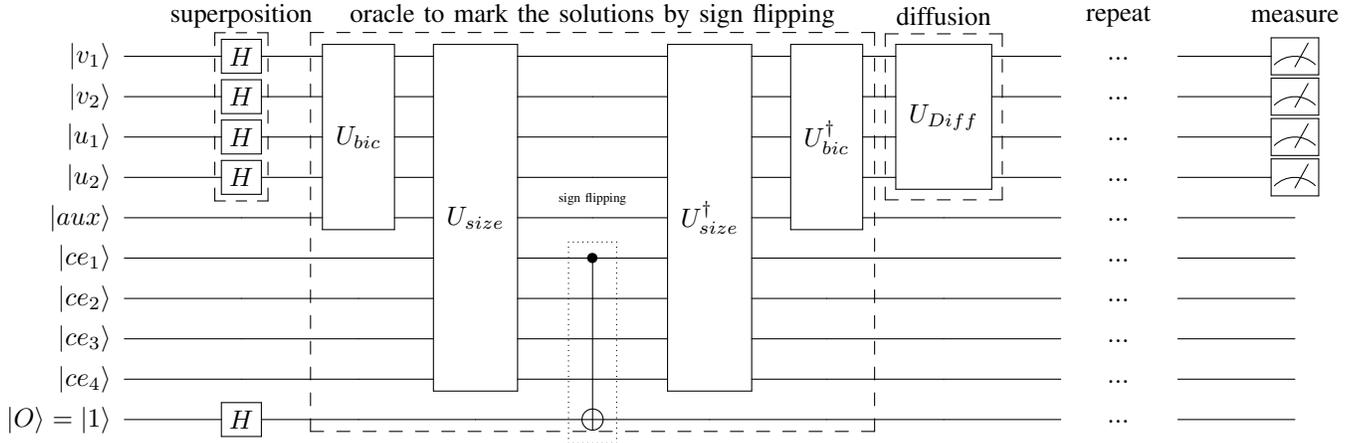

\subsection{Our Algorithm: qMBS}
Before proposing the final algorithm qMBS to find the maximum biclique, 
we first present a subprocedure qKBS that finds a size-$k$ biclique. 
Recall that at  Step 2  of the Grover's search, we need to flip the \textit{amplitude} of a size-$k$ biclique  state $\ket{x}$ (please refer to Figure~\ref{grover}a,~\ref{grover}b). 
For this purpose, 
we introduce an oracle qubit $\ket{O}$ which is initially set to be $\ket{1}$, 
and transforms to $(\ket{0}-\ket{1})/\sqrt 2$ after being operated by an $H$ gate. 
By now we have $\ket{x}\ket{ce_k}\ket{O}=\ket{x}\ket{1}(\ket{0}-\ket{1})/\sqrt 2$. 
We then use a CNOT gate to act on $\ket{ce_k}$ and $\ket{O}$, 
after which 
$\ket{O}$ transforms to $-(\ket{0}-\ket{1})/\sqrt 2$. 
Since the negative sign can be moved backward across the tensor product, 
 the  $\ket{x}\ket{ce_k}\ket{O}$ now transforms to $-\ket{x}\ket{ce_k}\ket{O}$. 
Therefore, the amplitude of $\ket{x}$ is flipped by a  negative sign. 
Now we can assemble the complete oracle with the Grover's search framework to find a size-$k$ biclique as the Quantum $k$-Biclique Search Algorithm (qKBS, shown in Algorithm~\ref{algo:qkbs}). 



\begin{algorithm}[tttt]
        \caption{Quantum $k$-Biclique Search: qKBS}
        \begin{algorithmic}[1] 
            \Require Graph $G(L,R,E)$, size $k$;
            \Ensure A biclique with size $k$ or $\emptyset$;
            \State Prepare the initial state to be an equal superposition of $2^n$ possible subsets of $L\cup R$;
            \State Use the oracle described in Section~\ref{subsec:bic}\&\ref{subsec:ec} with $\ket{O}$ to flip  the amplitude signs of the $k$-biclique states;
            \State Use a diffusion operator to inverse the amplitude of each base state about the amplitude average; 
            \State Repeat Line 2\&3 for $\lfloor\frac{\pi}{4}\sqrt{2^n/M}\rfloor$ times, then measure the final state of the $n$ vertex qubits;
			\State Output the $k$-biclique or $\emptyset$;
        \end{algorithmic}
    \label{algo:qkbs} 
   
    \end{algorithm}

\begin{algorithm}[tttt]
        \caption{Quantum Maximum Biclique Search: qMBS}
        \begin{algorithmic}[1] 
            \Require Graph $G(L,R,E)$;
            \Ensure A maximum biclique;
            \State Use  qKBS to search for a biclique with  size $k$, and find the maximum $k\in [1,m]$ by binary search; 
            \State Output the maximum biclique;

        \end{algorithmic}
    \label{algo:qmbs} 
   
    \end{algorithm}

Note that in Algorithm~\ref{algo:qkbs}, $M$ denotes the number of size-$k$ bicliques in the graph, which can be estimated by    the quantum counting algorithm~\cite{brassard1998quantum}. 
The circuit for searching   a size-$1$ biclique   is shown in Figure~\ref{full}. 
Here we use $\ket{aux}$ to summarize   the auxiliary qubits. 
$U_{bic}$ is the biclique checking circuit and $U_{size}$ is the edge counting circuit. 
Since we need to place all the auxiliary qubits to   their initial states after each iteration, the inverses  $U_{size}^{-1}=U_{size}^\dag$ and $U_{bic}^{-1} = U_{bic}^\dag$ are applied sequentially. 
Due to that the inverse of a CNOT gate is  itself, $U^\dag$ contains exactly the same gates as $U$ with  a  reverse order.  
Now we can present our  algorithm qMBS to search for a maximum biclique as the Quantum Maximum Biclique Search Algorithm (qMBS, shown in Algorithm~\ref{algo:qmbs}).  



   

\textbf{Resource requirement.}
Landauer's Principle~\cite{5392446} asserts that there is a minimum possible amount of energy required to erase one bit of information, known as the Landauer limit. This limit is given as $kT_k\ln(2)$, where $k$ is Boltzmann's constant  and $T_k$ is the temperature of the system. 
An normal computer continuously erases information during the computing process because its  underlying logic of computing is to use irreversible logic gates, 
 e.g., AND gate map two bits of information into one bit, 
 and is not possible to infer the two-bit input from the one-bit output. 
Due to the property of the unitary matrix 
$U^{-1} =U^\dag$,
all the quantum gates in   qMBS are reversible. 
This indicates that most of the computation units (except for the final measurements) require far less energy resource than normal algorithms according to Landauer's Principle.
Even though in practice  dissipation is required for system stability and immunity from noise, 
our algorithms are still promising for computation
on large-scale graphs in terms of energy consumption in near future. 


\subsection{Complexity Analysis}
The space complexity 
is quantified by the qubit number, and the time complexity 
is quantified by the gate number. 
According to the analysis in Section~\ref{subsec:bic}\&\ref{subsec:ec}, 
the qKBS and  qMBS have the same space complexity $O(n^2)$.
The number of CNOT gates in an oracle is $O(n^2)$  because $U$ and $U^\dag$ contain the same number of gates. 
The number of CNOT gates in the diffusion operator is $O(n)$~\cite{nielsen_chuang_2010}. 
The number of iterations of the oracle and the diffusion is $O(\sqrt{2^{n}})$. 
The number of $H$ gates for preparing the equal superposition    is $O(n)$. 
Therefore, the total time complexity of qKBS is 
$O(n +( n^2+n)\sqrt{2^n}) = O(n^2\sqrt{2^n})$, which is $O^*(\sqrt{2^n})$. 
The qMBS involves at most $O(\log m) = O(\log n)$  iterations of qKBS, so the total time complexity  is $O(n^2\log n\sqrt{2^n})$, which is also $O^*(\sqrt{2^n})$. 


\subsection{Maximum Vertex Biclique Problem and Maximum Balanced Biclique Problem: qMBSv and qMBSb}

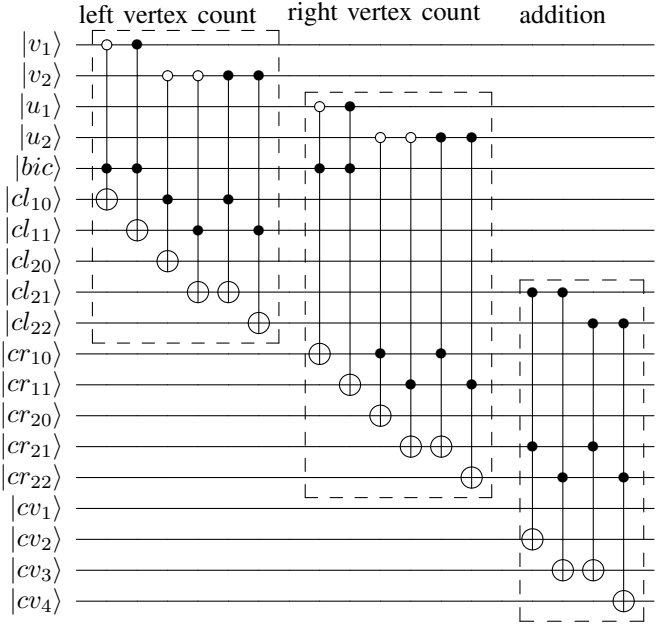
\begin{figure}[t]
\[
\Qcircuit @C=.33em @R=.35em @! {
& & & & \mbox{left vertex count}& & & & & & &\hspace{0.1cm}\mbox{right vertex count} & & & & & &\mbox{addition} & &  \\
&\lstick{\ket{v_1}} & \ctrlo{4} & \ctrl{4} & \qw & \qw & \qw & \qw & \qw & \qw & \qw & \qw & \qw & \qw & \qw & \qw & \qw & \qw & \qw & \qw & \qw\\
&\lstick{\ket{v_2}} & \qw & \qw & \ctrlo{4} & \ctrlo{5} & \ctrl{4} & \ctrl{5} & \qw & \qw & \qw & \qw & \qw & \qw & \qw & \qw & \qw & \qw & \qw & \qw & \qw\\
&\lstick{\ket{u_1}} & \qw & \qw & \qw & \qw  & \qw & \qw & \qw & \ctrlo{2} & \ctrl{2} & \qw & \qw & \qw & \qw & \qw & \qw & \qw & \qw & \qw & \qw\\
&\lstick{\ket{u_2}} & \qw & \qw & \qw & \qw  & \qw & \qw & \qw & \qw & \qw & \ctrlo{7} & \ctrlo{8} & \ctrl{7} & \ctrl{8} & \qw & \qw & \qw & \qw & \qw & \qw\\
&\lstick{\ket{bic}} & \ctrl{1} & \ctrl{2} & \qw & \qw  & \qw & \qw & \qw & \ctrl{6} & \ctrl{7} & \qw & \qw & \qw & \qw & \qw & \qw & \qw & \qw & \qw & \qw\\
&\lstick{\ket{cl_{10}}} & \targ & \qw & \ctrl{2} & \qw & \ctrl{3} & \qw & \qw & \qw & \qw & \qw & \qw & \qw & \qw & \qw & \qw & \qw & \qw & \qw & \qw\\
&\lstick{\ket{cl_{11}}} & \qw &\targ & \qw & \ctrl{2} & \qw & \ctrl{3} & \qw & \qw & \qw & \qw & \qw & \qw & \qw & \qw & \qw & \qw & \qw & \qw & \qw\\
&\lstick{\ket{cl_{20}}} & \qw & \qw & \targ & \qw & \qw & \qw & \qw & \qw & \qw & \qw & \qw & \qw & \qw & \qw & \qw & \qw & \qw & \qw & \qw\\
&\lstick{\ket{cl_{21}}} & \qw & \qw & \qw & \targ & \targ & \qw & \qw & \qw & \qw & \qw & \qw & \qw  & \qw & \qw & \ctrl{5} & \ctrl{6} & \qw & \qw & \qw\\
&\lstick{\ket{cl_{22}}} & \qw & \qw & \qw & \qw & \qw & \targ & \qw & \qw & \qw & \qw & \qw & \qw  & \qw & \qw & \qw & \qw & \ctrl{4} & \ctrl{5} & \qw\\
&\lstick{\ket{cr_{10}}} & \qw & \qw & \qw & \qw & \qw & \qw & \qw & \targ & \qw & \ctrl{2} & \qw &\ctrl{3}  & \qw & \qw & \qw & \qw & \qw & \qw & \qw\\
&\lstick{\ket{cr_{11}}} & \qw & \qw & \qw & \qw & \qw & \qw & \qw & \qw & \targ & \qw &\ctrl{2} & \qw & \ctrl{3} & \qw & \qw & \qw & \qw & \qw & \qw\\
&\lstick{\ket{cr_{20}}} & \qw & \qw & \qw & \qw & \qw & \qw & \qw & \qw & \qw & \targ & \qw & \qw & \qw & \qw & \qw & \qw & \qw & \qw & \qw \\
&\lstick{\ket{cr_{21}}} & \qw & \qw & \qw & \qw & \qw & \qw & \qw & \qw & \qw & \qw &\targ & \targ & \qw & \qw & \ctrl{3} & \qw & \ctrl{4} & \qw & \qw \\
&\lstick{\ket{cr_{22}}} & \qw & \qw & \qw & \qw & \qw & \qw & \qw & \qw & \qw & \qw & \qw & \qw & \targ & \qw & \qw & \ctrl{3} & \qw & \ctrl{4} & \qw \\
&\lstick{\ket{cv_{1}}}  & \qw & \qw & \qw & \qw & \qw & \qw & \qw & \qw & \qw & \qw & \qw & \qw & \qw & \qw & \qw & \qw & \qw & \qw & \qw \\
&\lstick{\ket{cv_{2}}}  & \qw & \qw & \qw & \qw & \qw & \qw & \qw & \qw & \qw & \qw & \qw & \qw & \qw & \qw & \targ & \qw & \qw & \qw & \qw \\
&\lstick{\ket{cv_{3}}}  & \qw & \qw & \qw & \qw & \qw & \qw & \qw & \qw & \qw & \qw & \qw & \qw & \qw & \qw & \qw & \targ & \targ & \qw & \qw \\
&\lstick{\ket{cv_{4}}}  & \qw & \qw & \qw & \qw & \qw & \qw & \qw & \qw & \qw & \qw & \qw & \qw & \qw & \qw & \qw & \qw & \qw & \targ & \qw 
\gategroup{2}{3}{11}{8}{.7em}{--}
\gategroup{4}{10}{16}{15}{.7em}{--}
\gategroup{10}{17}{20}{20}{.7em}{--}
}
\]
\caption{Vertex addition for maximum vertex biclique search}
\label{vcnt}
\end{figure}

\textbf{Maximum Vertex Biclique Problem.}
The maximum vertex biclique problem that searches for a biclique $C$ with the maximum number of vertices $|L(C)\cup R(C)|$. 
To propose a quantum algorithm to solve this problem, we only need to replace the multiplication mapping
 of the edge counting of qMBS 
 with an addition mapping to count vertices: 
 \begin{equation}
 	(\ket{cl_{|L|\mbox{ }i}}, \ket{cr_{|R|\mbox{ }j}}) \mapsto \ket{cv_{i+j}}
 \end{equation}
Figure~\ref{vcnt} shows the quantum circuit. 
The only difference between Figure~\ref{ecnt} and Figure~\ref{vcnt} is the third dashed box, where we use the addition mapping to replace the multiplication mapping $(\ket{cl_{|L|\mbox{ }i}}, \ket{cr_{|R|\mbox{ }j}}) \mapsto \ket{ce_{i\cdot j}}$. 
We name the algorithm to find a maximum vertex biclique as qMBSv. 

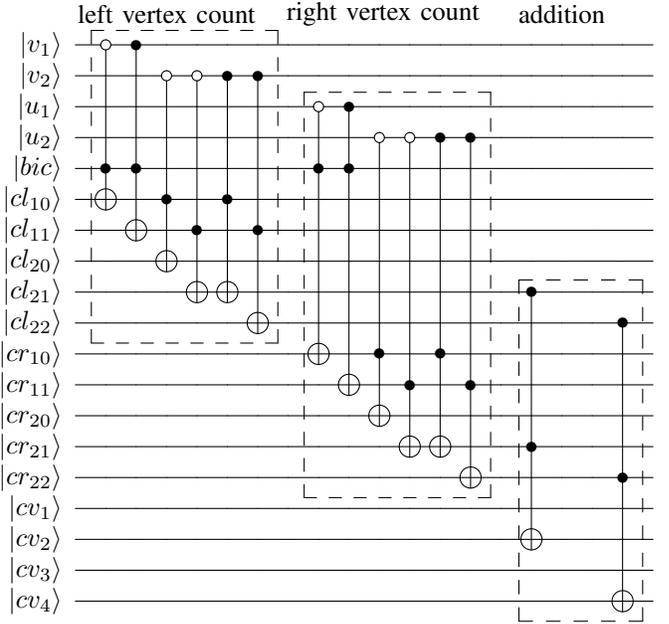
\begin{figure}[t]
\[
\Qcircuit @C=.33em @R=.35em @! {
& & & & \mbox{left vertex count}& & & & & & &\hspace{0.1cm}\mbox{right vertex count} & & & & & &\mbox{addition} & &  \\
&\lstick{\ket{v_1}} & \ctrlo{4} & \ctrl{4} & \qw & \qw & \qw & \qw & \qw & \qw & \qw & \qw & \qw & \qw & \qw & \qw & \qw & \qw & \qw & \qw & \qw\\
&\lstick{\ket{v_2}} & \qw & \qw & \ctrlo{4} & \ctrlo{5} & \ctrl{4} & \ctrl{5} & \qw & \qw & \qw & \qw & \qw & \qw & \qw & \qw & \qw & \qw & \qw & \qw & \qw\\
&\lstick{\ket{u_1}} & \qw & \qw & \qw & \qw  & \qw & \qw & \qw & \ctrlo{2} & \ctrl{2} & \qw & \qw & \qw & \qw & \qw & \qw & \qw & \qw & \qw & \qw\\
&\lstick{\ket{u_2}} & \qw & \qw & \qw & \qw  & \qw & \qw & \qw & \qw & \qw & \ctrlo{7} & \ctrlo{8} & \ctrl{7} & \ctrl{8} & \qw & \qw & \qw & \qw & \qw & \qw\\
&\lstick{\ket{bic}} & \ctrl{1} & \ctrl{2} & \qw & \qw  & \qw & \qw & \qw & \ctrl{6} & \ctrl{7} & \qw & \qw & \qw & \qw & \qw & \qw & \qw & \qw & \qw & \qw\\
&\lstick{\ket{cl_{10}}} & \targ & \qw & \ctrl{2} & \qw & \ctrl{3} & \qw & \qw & \qw & \qw & \qw & \qw & \qw & \qw & \qw & \qw & \qw & \qw & \qw & \qw\\
&\lstick{\ket{cl_{11}}} & \qw &\targ & \qw & \ctrl{2} & \qw & \ctrl{3} & \qw & \qw & \qw & \qw & \qw & \qw & \qw & \qw & \qw & \qw & \qw & \qw & \qw\\
&\lstick{\ket{cl_{20}}} & \qw & \qw & \targ & \qw & \qw & \qw & \qw & \qw & \qw & \qw & \qw & \qw & \qw & \qw & \qw & \qw & \qw & \qw & \qw\\
&\lstick{\ket{cl_{21}}} & \qw & \qw & \qw & \targ & \targ & \qw & \qw & \qw & \qw & \qw & \qw & \qw  & \qw & \qw & \ctrl{5} & \qw & \qw & \qw & \qw\\
&\lstick{\ket{cl_{22}}} & \qw & \qw & \qw & \qw & \qw & \targ & \qw & \qw & \qw & \qw & \qw & \qw  & \qw & \qw & \qw & \qw & \qw & \ctrl{5} & \qw\\
&\lstick{\ket{cr_{10}}} & \qw & \qw & \qw & \qw & \qw & \qw & \qw & \targ & \qw & \ctrl{2} & \qw &\ctrl{3}  & \qw & \qw & \qw & \qw & \qw & \qw & \qw\\
&\lstick{\ket{cr_{11}}} & \qw & \qw & \qw & \qw & \qw & \qw & \qw & \qw & \targ & \qw &\ctrl{2} & \qw & \ctrl{3} & \qw & \qw & \qw & \qw & \qw & \qw\\
&\lstick{\ket{cr_{20}}} & \qw & \qw & \qw & \qw & \qw & \qw & \qw & \qw & \qw & \targ & \qw & \qw & \qw & \qw & \qw & \qw & \qw & \qw & \qw \\
&\lstick{\ket{cr_{21}}} & \qw & \qw & \qw & \qw & \qw & \qw & \qw & \qw & \qw & \qw &\targ & \targ & \qw & \qw & \ctrl{3} & \qw & \qw & \qw & \qw \\
&\lstick{\ket{cr_{22}}} & \qw & \qw & \qw & \qw & \qw & \qw & \qw & \qw & \qw & \qw & \qw & \qw & \targ & \qw & \qw & \qw & \qw & \ctrl{4} & \qw \\
&\lstick{\ket{cv_{1}}}  & \qw & \qw & \qw & \qw & \qw & \qw & \qw & \qw & \qw & \qw & \qw & \qw & \qw & \qw & \qw & \qw & \qw & \qw & \qw \\
&\lstick{\ket{cv_{2}}}  & \qw & \qw & \qw & \qw & \qw & \qw & \qw & \qw & \qw & \qw & \qw & \qw & \qw & \qw & \targ & \qw & \qw & \qw & \qw \\
&\lstick{\ket{cv_{3}}}  & \qw & \qw & \qw & \qw & \qw & \qw & \qw & \qw & \qw & \qw & \qw & \qw & \qw & \qw & \qw & \qw & \qw & \qw & \qw \\
&\lstick{\ket{cv_{4}}}  & \qw & \qw & \qw & \qw & \qw & \qw & \qw & \qw & \qw & \qw & \qw & \qw & \qw & \qw & \qw & \qw & \qw & \targ & \qw 
\gategroup{2}{3}{11}{8}{.7em}{--}
\gategroup{4}{10}{16}{15}{.7em}{--}
\gategroup{10}{17}{20}{20}{.7em}{--}
}
\]
\caption{Vertex addition for maximum balanced biclique search}
\label{vcnt2}
\end{figure}

\textbf{Maximum Balanced Biclique Problem.}
The maximum balanced biclique problem  searches for a maximum vertex biclique $C$ with $|L(C)| = |R(C)|$. 
To propose a quantum variants of qMBSv for this problem,  we only need to restrict the addition mapping with a condition that $i=j$. 
Figure~\ref{vcnt2} shows the quantum circuit. 
The only difference between Figure~\ref{vcnt} and Figure~\ref{vcnt2} is the third dashed box, where we restrict that the addition can only be performed for two equal numbers, e.g., $i=j=1$ or $i=j=2$. 
We name the algorithm to find a maximum balanced biclique as qMBSb. 
It can be verified that the time and space complexities of qMBSv and qMBSb are same as qMBS. 
\section{Experimental Studies}\label{sec:exp}

In this section, we conduct proof-of-principle  experiments   using state-of-the-art IBM quantum simulators. 
The experiment is divided into two parts: 
\begin{itemize}
	\item [I.] We test our algorithms for the example graph in Figure~\ref{bic1}, 
the size of which is comparable to existing works on quantum circuits  for clique problems (details in Table~\ref{datasets}). 
To provide a comprehensive discussion about the algorithm behaviors
of searching
bicliques with all possible sizes, instead of a binary search, we implement qMBS by 
calling qKBS sequentially from $k=1$ to $k=m$. 
\item [II.] We compare qMBS with the state-of-the-art~\cite{lyu2022maximum} across 10 synthetic datasets, where the number of vertices in the datasets ranges from 6 to 10, 
and  the number of edges ranges from 3 to 23. 
The graph size is significantly larger than that of existing quantum graph database works~\cite{chang2018quantum,metwalli2020finding} (details in Table~\ref{datasets}). 
For fairness, we utilize the complete binary search version of the qMBS algorithm. 
\end{itemize}
All the experiments are conducted in Python 3.8 with Qiskit  and tested on  IBM  simulators (details in Table~\ref{simulators}).  

\begin{table}
\renewcommand\arraystretch{1.2}
\centering
\caption{Comparison of dataset sizes}
\begin{tabular}{p{76pt}p{123pt}p{2pt}p{2pt}}
\toprule
Problem  & Time complexity \& Work &$n$ & $m$      \\
\midrule
Maximum clique & $O^*(2^{\frac{n}{2}})$~\cite{chang2018quantum}&  2  &  4  \\
$k$-clique  & $O^*(2^{\frac{n}{2}})$~\cite{metwalli2020finding}&  4 &   4   \\
Maximum biclique  &$O^*(2^{\frac{n}{2}})$ [qMBS] &  10 &   22   \\
\bottomrule
\end{tabular}
\label{datasets}
\end{table}

\begin{figure*}[t]
	\centering
	\subfloat[State distribution of qMBS \label{9a}]{\includegraphics[width=7.75cm]{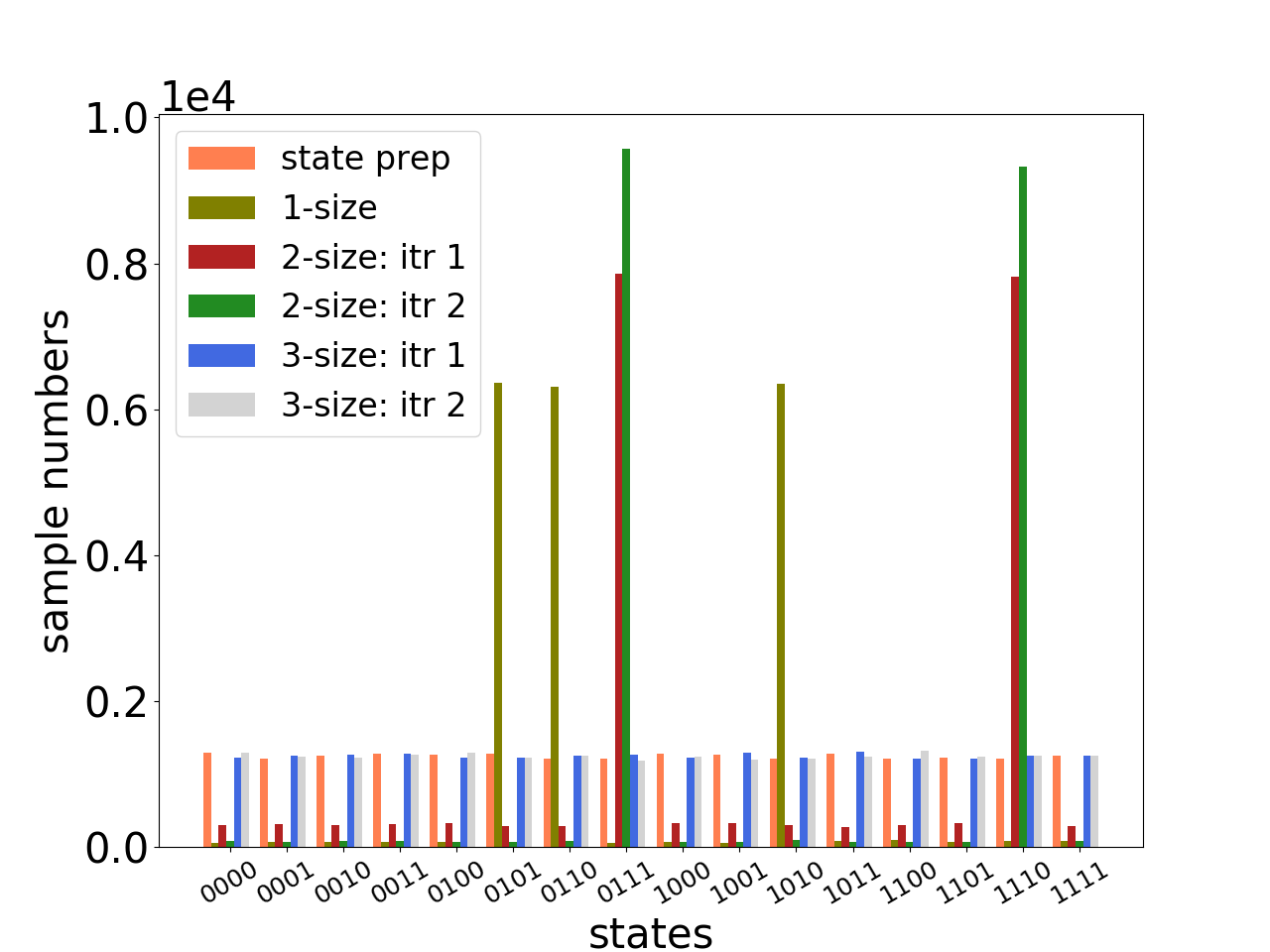}}
	\subfloat[State distribution of qMBSv \label{9b}]{\includegraphics[width=7.75cm]{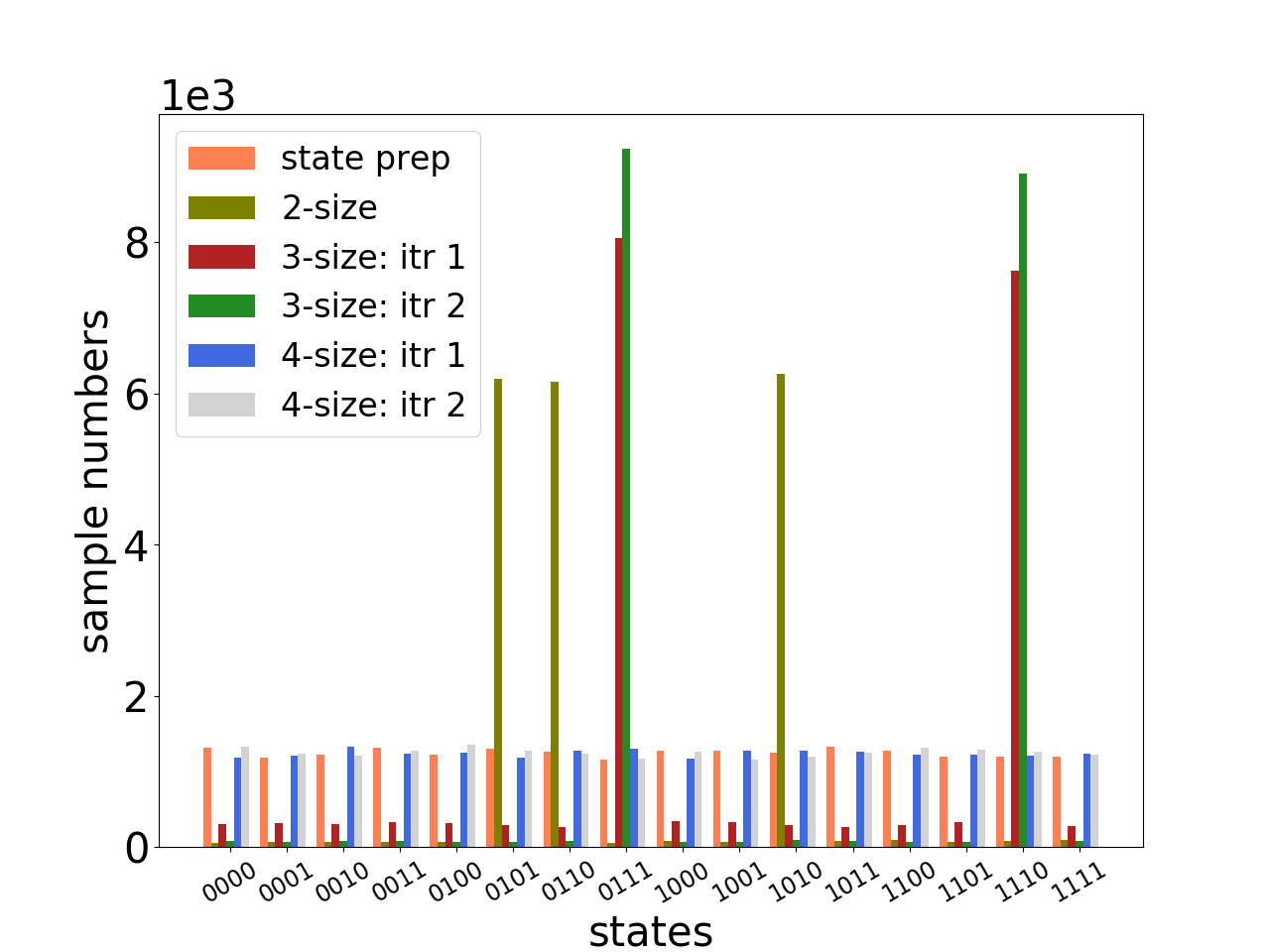}}	
	
	\subfloat[State distribution of qMBSb \label{9c}]{\includegraphics[width=7.75cm]{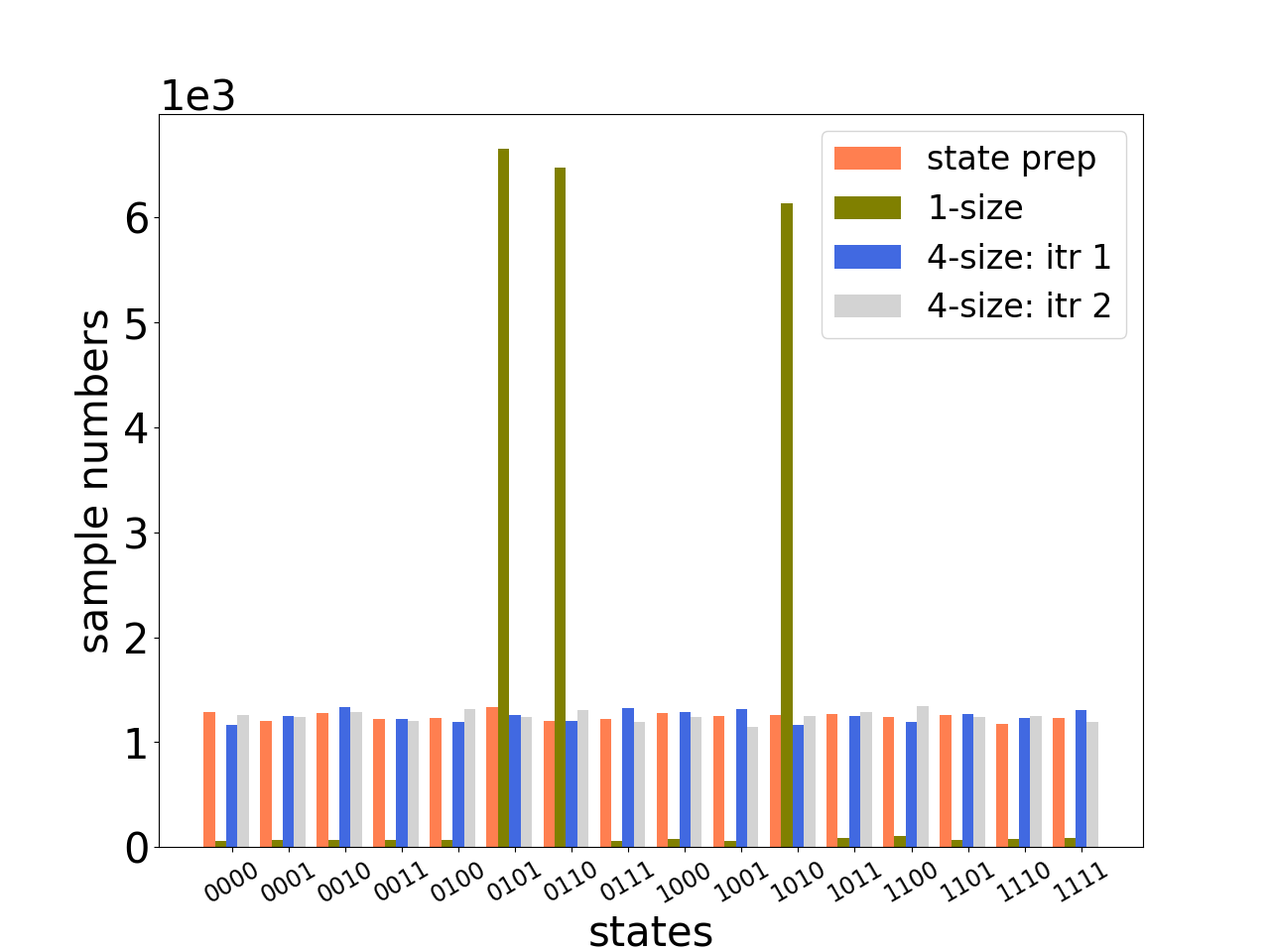}}
\subfloat[Running time  \label{9d}]{\includegraphics[width=7.75cm]{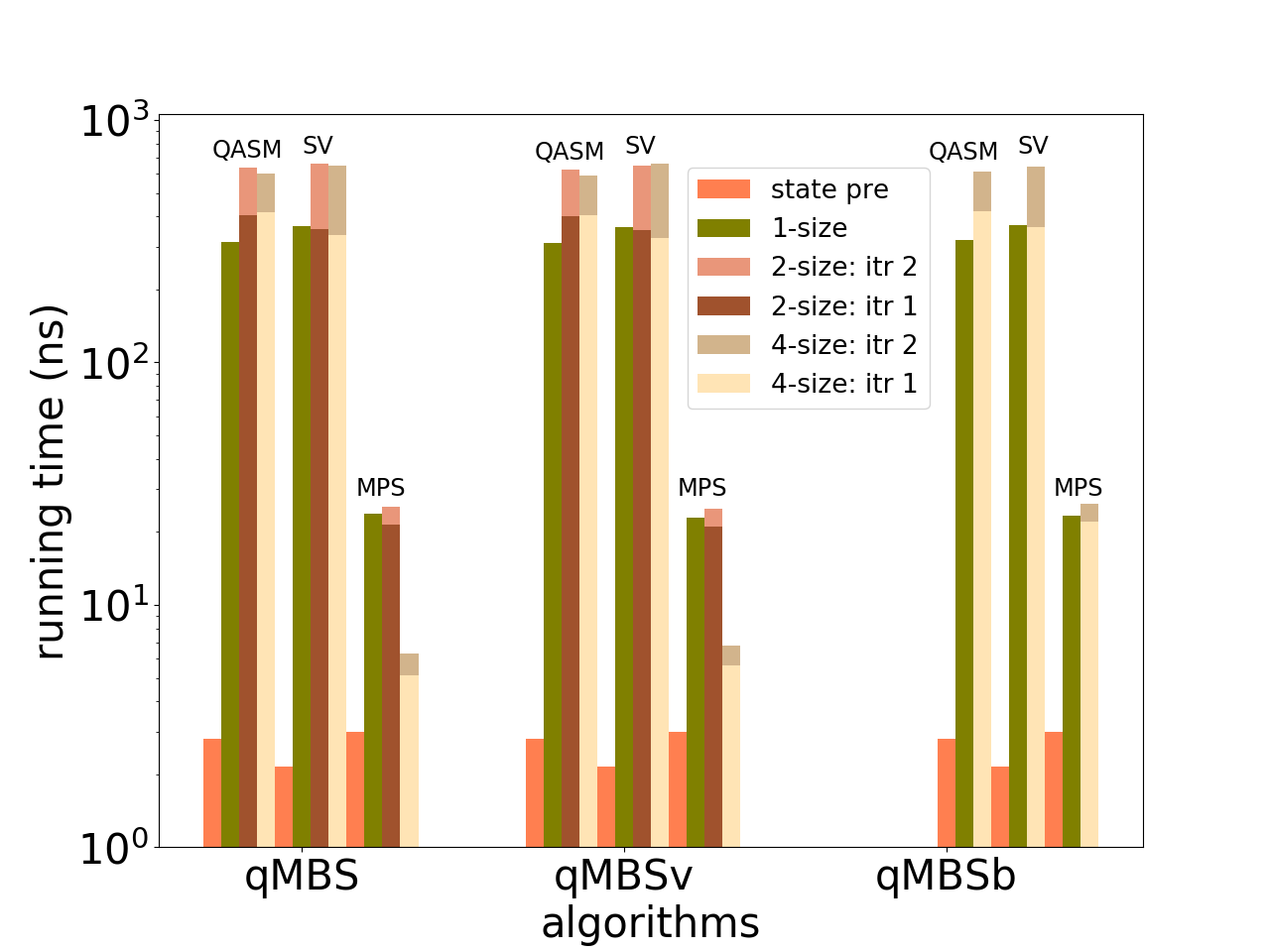}}	
\vspace{5pt}
	\captionsetup{justification=centering}
	\caption{State distribution and running time}\label{fig:exp}
	
\end{figure*}

\subsection{Error probability convergence}
Given the inherent indeterminacy of quantum computing, there exists a probability of error whereby, upon measurement, the final state incorrectly collapses into a non-solution state. This inherent indeterminacy is a fundamental characteristic of quantum computing and cannot be theoretically eradicated. 
The error probability is at most $\pi^2/(4T)^2$, where $T$ denotes the number of iterations~\cite{nielsen_chuang_2010}. For a relatively low value of $T$, we can execute the qKBS algorithm multiple times, e.g.,  $c$ times. This approach reduces the error probability to $\pi^2/(4T)^{2c}$. As such, the error rate is anticipated to rapidly diminish to a level that is significantly lower than the thermal noise inherent in physical devices. This allows our algorithms to be safely employed in practical settings to procure precise solutions.
To evaluate the practical error rate, we execute our algorithms with 20K shots and measure the final states to report the frequency distribution across 16 possible base states (ranging from $\ket{0000}$ to $\ket{1111}$). All the algorithms undergo testing on three simulators. However, due to the high similarity in distributions, we only present the results obtained from the QASM simulator.

\begin{table}
\renewcommand\arraystretch{1.2}
\centering
\caption{Simulators}
\begin{tabular}{l l l }
\toprule
Simulator  & Qubits & Type      \\
\midrule
QASM  &  32  &  General, context-aware  \\
Statevector   &  32 &   Schr\"{o}dinger wavefunction \\
MPS   &  100 &   Matrix product state   \\
\bottomrule
\end{tabular}
\label{simulators}
\end{table}

Figure~\ref{9a} presents the results of the qMBS algorithm. After state preparation, the distribution of base states generally appears uniform, hence it is referred to as an equal superposition.
To locate a biclique of size 1, we proceed with a single iteration of Steps 2 and 3 in the qKBS algorithm. The results yield three significant peaks at $\ket{0101}$, $\ket{0110}$, and $\ket{1010}$. These peaks correspond to the three bicliques of size 1, namely $\{v_2, u_2\}$, $\{v_2, u_1\}$, and $\{v_1, u_1\}$. The probability of error, defined by the final state not collapsing into one of these three peaks, is calculated to be 4.87\%. This value is significantly lower than the theoretically guaranteed error rate of $\pi^2/(4T)^2$.
To discover a biclique of size 2, two iterations are required. Interestingly, after the first iteration (denoted as $itr\mbox{ }1$ in Figure~\ref{9a}), two prominent peaks can be observed at $\ket{0111}$ and $\ket{1110}$, corresponding to the bicliques ${v_2,u_1,u_2}$ and ${v_1,v_2,u_1}$, respectively. If we measure the state at this point, the associated error rate stands at 21.59\%. 
%
%
Upon the completion of the second iteration, the peaks become more pronounced and the error rate is significantly reduced to 5.53\%. When endeavoring to find a biclique of size 3, the qKBS algorithm encounters difficulties as the oracle fails to mark a solution state. Consequently, the diffusion operator lacks a specific target to amplify, resulting in the states following a uniform distribution after two iterations. Ultimately, the qMBS algorithm identifies a biclique of size 2 as the optimal solution.
In the case of the qMBSv algorithm, the results presented in Figure~\ref{9b} closely resemble those of Figure~\ref{9a}. This is because a biclique with one edge inherently corresponds to a biclique with two vertices.
As for the qMBSb algorithm depicted in Figure~\ref{9c}, given that the number of edges in a balanced biclique can only be a square number, there is no need to search for bicliques of sizes 2 and 3.
The error probabilities are generally around 5\% for these small instances, which provides a practical effectiveness guarantee (that decreases proportional to $1/T^2$) for larger datasets in future applications. This indicates the robustness of our quantum algorithms, promising significant potential for tackling larger and more complex problems.

\begin{table*}
\renewcommand\arraystretch{1.2}
\centering
\caption{Comparison with state-of-the-art}
\begin{tabular}{l l l l l l l l l l l}
\toprule
Dataset  & $D_{6,3}$ & $D_{6,6}$ &$D_{7,6}$  &$D_{7,11}$&$D_{8,5}$&$D_{8,14}$&$D_{9,4}$&$D_{9,18}$&$D_{10,7}$&$D_{10,23}$   \\
\midrule
Maximum biclique size&2&4&4&9&3&12&3&16&4&20\\
Running time of MBC$^*$ (ns)&573.3&563.4&583.5&581.5&627.3&639.6&803.2&807,9&925.8&931.6\\
Running time of qMBS (ns)&43.3&44.2&49.5&54.3&62.7&64.4&68.5&68.2&77.9&79.5\\
Error probability&$<10^{-2}$&$<10^{-2}$&$<10^{-3}$&$<10^{-3}$&$<10^{-3}$&$<10^{-3}$&$<10^{-4}$&$<10^{-4}$&$<10^{-4}$&$<10^{-4}$\\

\bottomrule
\end{tabular}
\label{comp}
\end{table*}

\subsection{Efficiency}

We evaluate the performance of all algorithms on three simulators, where the reported running time is calculated as an average over 20,000 executions. The results are presented in Figure~\ref{9d}.
Focusing initially on the results from the QASM simulator, we observe the following. For the qMBS algorithm, the state preparation phase is completed in a swift 2.8 nanoseconds (ns), a duration that is negligible in comparison to the time required for the subsequent iterations. The first iteration, which is aimed at identifying a biclique of size 1, requires 315ns. Subsequently, the first and second iterations to search for a biclique of size 2 consume 405ns and 230ns respectively. The cumulative duration for the two rounds of iterations targeted at identifying a biclique of size 3 amounts to 600ns. The entirety of the qMBS algorithm run therefore requires approximately 1500ns (1200ns), 
where the values in parentheses show the time if binary search applies. 
The performance of the qMBSv algorithm closely mirrors that of the qMBS algorithm due to the similar structure and operations they share. On the other hand, the qMBSb algorithm, despite each iteration consuming a similar duration to those in the qMBS and qMBSv algorithms, has a noticeably fewer number of iterations. This is because it only needs to consider bicliques whose sizes are square numbers, thereby reducing the computational complexity. Consequently, the running time of the qMBSb algorithm is approximately 900ns (900ns).


The results obtained from the Statevector simulator closely parallel those of QASM. However, a remarkable speed-up is evident when using the MPS (Matrix Product State) simulator, which owes its efficiency to the effective representation of matrix product states.
For the qMBS algorithm, the MPS simulator spends a mere 3ns preparing the initial equal superposition. The identification of a 1-size biclique takes 23.8ns. The two iterations that target a 2-size biclique are completed in 21.5ns and 4ns respectively, whereas the two iterations aimed at discovering a 3-size biclique consume only 5.1ns and 0.8ns. In total, the entire running time is about 55ns (or 30ns if binary search is applied) for qMBS. The qMBSv and qMBSb algorithms register similar timings at 55ns (or 30ns with binary search) and 30ns respectively. Here, the values in parentheses denote the results obtained using the binary search strategy.
The significant speed-up observed on the MPS simulator suggests that our algorithms generate states with low levels of entanglement, which the matrix product state representation can simulate efficiently.

\subsection{Comparison with state-of-the-art}
Due to the limitations of existing hardware, even though our algorithm outperforms the state-of-the-art in terms of complexity and resource consumption, 
large-scale QPUs are not yet prepared  
to test the algorithm on large datasets. 
Nevertheless, we still aspire to compare our algorithm with the state-of-the-art on small datasets. Given that the largest quantum simulator available to us currently supports up to 100 qubits, our algorithm can be tested on bipartite graphs of about 10 vertices with the MPS simulator. To make the test results more generally meaningful, we have examined a total of 10 synthetic datasets with vertex counts ranging from 6 to 10. We denote a dataset as $D_{i,j}$, where $i$ represents the vertex number of the dataset and $j$ represents the edge number of the dataset. For each identical size $i$, we selected two different $j$ values, one small and one large, to ensure that the experiment covers both small and large biclique situations. The datasets and experimental results are shown in Table~\ref{comp}. The reported running time is calculated as an average over 20,000 executions. 

	We observe that across all datasets, qMBS is approximately an order of magnitude faster than MBC$^*$. The efficiency of qMBS is affected by both the number of vertices and the number of edges in the dataset. 
	As the dataset size grows, the increase in running time is slower compared to MBC$^*$, which is a result of the efficiency boost brought about by the quadratic speed-up of qMBS in terms of time complexity. With the increase in the number of iterations, the error probability decreases exponentially. For a graph with 10 vertices, the error probability is already less than $10^{-4}$. Therefore, when actually applied to large-scale datasets, this error probability is generally lower than the thermodynamic noise of the device and can be neglected.

\subsection{Summary}
In summary, the experimental results 
underscore the proficiency of our proposed algorithms. They quickly evolve the initial state into the solution state, and maintain the error probability at a negligible level even with small iteration numbers. 
Compared to the state-of-the-art method~\cite{lyu2022maximum}, qMBS demonstrates an order-of-magnitude improvement in efficiency on small datasets, and the growth rate of its running time is slower than state-of-the-art methods as the size of the graph increases.
This underlines the practical potential of our proposed algorithms in quantum computation, promising rapid and accurate solution-finding 
in future.

\section{Related Works}\label{sec:related}
Related works can be categorized into two types: those on biclique-related problems, and those on quantum database or graph database algorithms.

\textbf{Biclique problems.}
There are mainly four types of biclique  problems. 
\textbf{Maximal biclique enumeration} finds all the maximal bicliques within a bipartite graph. 
A time delay algorithm was proposed by the work~\cite{li2007maximal}, 
aiming to strike a balance between computational efficiency and resource usage.  
The work ~\cite{zhang2014finding} combined backtracking with a branch-and-bound framework to filter out unpromising search branches. 
Parallel algorithms with shared memory were designed by~\cite{8990406}. 
Pivot-based algorithms with index and batch-pivots-based algorithm for sparse bipartite graphs were proposed by \cite{abidi2020pivot} and \cite{chen2022efficient} respectively. 
\textbf{Maximum vertex biclique search} 
studies the problem of  finding a biclique with maximum number of vertices, which 
is polynomially solvable~\cite{lewis1983michael}. 
This problem was solved by formulating it as an instance of integer linear programming (ILP)~\cite{dawande2001bipartite},   
or by  being reduced to finding maximum flow in a constructed flow network~\cite{konig1931grafok}.
\textbf{Maximum edge biclique search} was proved to be NP-hard~\cite{peeters2003maximum}. 
An ILP solver was proposed 
by the work~\cite{sozdinler2018finding}. 
The work~\cite{lyu2022maximum} proposed a progressive-bounding framework 
for  large graphs. 
A probabilistic algorithm  using a Monte Carlo subspace clustering approach was designed by~\cite{shaham2016finding}. 
The work~\cite{feng2018parameterized} studied the parameterized maximum biclique problem  that determines if there exists a biclique with at least a given number of edges. 
Beside, the work~\cite{shahinpour2017scale} solved this problem by ILP on a general graph. 
The problem of \textbf{maximum balanced biclique search} looks for a maximum edge/vertex biclique $C$ with $|L(C)| = |R(C)|$. 
The work~\cite{mccreesh2014exact} proposed a branch and bound approach with symmetry breaking technique,  
based on which the work~\cite{zhou2018towards} designed an upper bound estimation method for further branch pruning. 
Algorithms for dense and sparse bipartite graphs were proposed  by \cite{chen2021efficient}. 
Besides, heuristic approaches 
were also studied by~\cite{tahoori2006application,al2007defect,wu2015review,wang2018new,zhou2019tabu,li2020general}. 
All of the aforementioned problems (except for the maximal biclique enumeration) can be solved in a quantum manner 
by qMBS and its variants. 
There are still some variants of the maximum biclique problem, e.g., personalized maximum biclique search~\cite{wang2022efficient}, maximal balanced signed biclique enumeration~\cite{sun2022maximal}, and vertex coverage for top-$k$ bicliques~\cite{abidi2022maximising}. 
Given the generality of the bipartite graph encoding proposed in our work, these  biclique tasks will also benefit from this encoding method, thereby facilitating researchers to propose corresponding quantum algorithms in the future.

\textbf{Quantum  database algorithms.}
There has been a recent surge of work concerning quantum database algorithms.
The  work\cite{10.14778/3598581.3598603} explored the transformative impact that quantum algorithms may have on the field of databases in both the immediate and near future.
The problem of multiple query optimization was studied on adiabatic quantum annealer by the work~\cite{10.14778/2947618.2947621}. 
The work~\cite{schonberger2022applicability} proposed circuit-based quantum algorithms for join order optimization,
based on which a variational quantum circuit~\cite{winker2023quantum} and a quantum annealer algorithm~\cite{nayak2023constructing} were proposed for this problem. 
Quantum computing has also invigorated research in  graph databases. Quantum walks have been employed for graph traversal~\cite{childs2004spatial}, and quantum PageRank algorithms show potential advantages over classical methods~\cite{paparo2013google}. Hardware like D-Wave's quantum annealing machines are tackling graph problems~\cite{mcgeoch2013experimental}, and quantum machine learning algorithms aim to leverage potential quantum benefits for graph data~\cite{schuld2021quantum,bai2021learning}.
Quantum algorithms have also been 
studied for 
clique problems, where a clique is a complete subgraph of a general graph. 
For the \textbf{maximum clique problem}, 
the work~\cite{bojic2012quantum} studied an oracle-based Grover's search, 
after which a concrete quantum circuit 
was designed by~\cite{chang2018quantum}. 
Different computation models were also studied for the problem, e.g., 
quantum adiabatic evolution~\cite{childs2002finding} and quantum annealing~\cite{chapuis2019finding}. 
However, these models are typically problem-specific and not as flexible as the quantum circuit when being generalized for other problems. 
For the \textbf{$\bm{k}$-clique problem},
the work~\cite{childs2005quantum} utilized quantum subset finding algorithms 
to find a size-$k$ clique with a small $k$. 
Grover's search based algorithms were also studied by \cite{metwalli2020finding}. 
These works cannot be applied to the biclique problems 
due to the bipartition restriction. 
To the best of our knowledge,
our work is the first to study a quantum approach for the biclique problems.
For all the quantum algorithms mentioned above, which are based on the quantum circuit computational model for the clique problem, it is worth noting that, while they are currently restricted by hardware limitations and not yet applicable on large-scale datasets, the theoretical acceleration in algorithmic complexity, coupled with the rapid advancement of quantum hardware in recent years, fosters hope that these quantum algorithms will outperform classical ones in real-world applications on large datasets in the near future. 




\vspace{10pt}
\section{Conclusion and Future Works}\label{sec:conclu}
In this work, 
we explored the potential of utilizing QPUs to expedite graph database algorithms, and have proposed a class of biclique problem algorithms based on quantum circuits. 
Specifically, we delved into the Maximum Biclique Problem (MBP) from a quantum perspective. A novel reversible quantum circuit was conceived for the purpose of determining whether a given subgraph constitutes a biclique of a certain size. Utilizing this, we introduced a quantum algorithm, qMBS, designed to address the MBP with a time complexity of $O^*(2^{\frac{n}{2}})$. Remarkably, this presents a quadratic acceleration compared to the state-of-the-art in terms of time complexity.
Furthermore, we elaborated on two extensions of qMBS to solve the Maximum Vertex Biclique problem and Maximum Balanced Biclique problem, broadening its applicability. To assess the practical performance of our proposed solutions, we conducted proof-of-principle experiments using state-of-the-art quantum simulators. These experimental results provided a substantial validation of our approach to the extent possible to date. 
The incorporation of reversible computing in our algorithms enhances their potential to handle real-world datasets in an energy-efficient manner, which adds significant value, considering the increasing importance of sustainability in computing. As quantum hardware continues to evolve, we anticipate our proposed algorithms will contribute to quantum computing's capability to tackle challenging  problems efficiently in the near future.

Future works will pivot towards another vital class of problems in graph databases: enumeration problems, such as maximal biclique/clique enumeration. In the context of NP-hard problem complexities, a search space of exponentially large branches can be perfectly accommodated in a superposition state within the $2^n$-dimensional space spanned by $n$ qubits. As a result, harnessing quantum algorithms to expedite enumeration problems in graph databases will constitute a significant direction for upcoming endeavors.

\newpage

\bibliographystyle{IEEEtran}  
\bibliography{IEEEabrv,IEEEexample}

\end{document}